\documentclass[fleqn,usenatbib]{mnras}

\usepackage{newtxtext,newtxmath}

\usepackage[T1]{fontenc}

\DeclareRobustCommand{\VAN}[3]{#2}
\let\VANthebibliography\thebibliography
\def\thebibliography{\DeclareRobustCommand{\VAN}[3]{##3}\VANthebibliography}

\usepackage{graphicx}	
\usepackage{amsmath}	
\usepackage{physics}

\def\be{\begin{equation}}
\def\ee{\end{equation}}

\newcommand{\msun}{\ensuremath{M_{\odot}}}
\newcommand{\mbh}{\ensuremath{M_\bullet}}

\title[QPE Timing for SMBH Mass Measurement]{Dynamical Measurement of Supermassive Black Hole Masses: QPE Timing Method}

\author[C. Zhou et al.]{
Cong Zhou,$^{1,2}$\thanks{E-mail: dysania@mail.ustc.edu.cn}
Zhen Pan,$^{3,4}$\thanks{E-mail: zhpan@sjtu.edu.cn}
Ning Jiang,$^{1,2}$\thanks{E-mail: jnac@ustc.edu.cn}
and Wen Zhao$^{1,2}$\thanks{E-mail: wzhao7@ustc.edu.cn}
\\
$^{1}$Department of Astronomy, University of Science and Technology of China, Hefei 230026, People’s Republic of China\\
$^{2}$School of Astronomy and Space Sciences, University of Science and Technology of China, Hefei 230026, People’s Republic of China\\
$^{3}$Tsung-Dao Lee Institute, Shanghai Jiao-Tong University, Shanghai, 520 Shengrong Road, 201210, People’s Republic of China\\
$^{4}$School of Physics \& Astronomy, Shanghai Jiao-Tong University, Shanghai, 800 Dongchuan Road, 200240, People’s Republic of China
}

\date{Accepted XXX. Received YYY; in original form ZZZ}

\pubyear{\the\year{}}

\begin{document}
\label{firstpage}
\pagerange{\pageref{firstpage}--\pageref{lastpage}}
\maketitle

\begin{abstract}
Quasi-periodic eruptions (QPEs) are intense repeating soft X-ray bursts with recurrence times about a few
hours to a few weeks from galactic nuclei. More and more analyses show that (at least a fraction of) QPEs are the result of collisions
between a stellar mass object (SMO, a stellar mass black hole or a main sequence star) and an accretion disk
around a supermassive black hole (SMBH) in galactic nuclei.  
Previous studies have shown  the possibility of reconstructing the SMO trajectory from QPE timing data, consequently 
 measuring the SMBH mass  from tracing a single SMO. 
In this paper, we construct a comprehensive Bayesian framework for implementing the QPE timing method, 
explore the optimal QPE observation strategy for measuring  SMBH masses,
and forecast the measurement precision expected in the era of multi-target X-ray telescope, Chasing All Transients Constellation Hunters (CATCH). Simulations of CATCH observations of GSN 069 and eRO-QPE2 like QPEs confirm the possible applications of 
 the QPE timing method  in precise measurement of SMBH masses (and spins), 
especially in the lower mass end ($\lesssim 10^7 M_\odot$) where QPEs prevail and relevant dynamical timescales are reasonably short to be measured. 
\end{abstract}




\section{Introduction} \label{sec:intro}

It is nowadays widely accepted that supermassive black holes (SMBHs) with masses ranging from $\sim 10^5-10^{10}$~\msun\ reside at the centers of most, if not all, massive galaxies with substantial spheroidal (bulge) components, both quiescent and active~(see reviews by \citealt{Kormendy1995,KH2013}). However, a precise measurement of the the masses of SMBHs (\mbh) has been a challenge since it is of great difficulty to directly probe the motion of objects in the vicinity of SMBHs, where the gravitational potential of the SMBHs exerts a dominant influence. In fact, aside from Sagittarius $\rm A^{\star}$ at the center of our Milky Way~\citep{Schodel2002,Ghez2005,Gillessen2009,GRAVITY2018b}, we have not been able to effectively resolve individual stars or gas clouds around SMBHs. Instead, we can only measure the collective motion of stars or gas within a given region as a whole even taking advantage of highly spatially resolved observations of Hubble Space Telescope (HST) and large adaptive optics (AO)-assisted ground-based telescopes. Therefore, the \mbh\ measurement suffers from additional uncertainties from velocity smoothing and galaxy central mass distribution in the dynamical modeling. So far, only a total of hundreds of SMBHs in the local galaxies ($\lesssim100$~Mpc) have been weighted by the so-called stellar and gas kinematics. Moreover, these SMBHs are found to be correlated with the bulge properties of their host galaxies, most commonly known as the \mbh-$\sigma_{\star}$ relation in which the $\sigma_{\star}$ refers to the stellar velocity dispersion~\citep{Ferrarese2000,Gebhardt2000}, making the field of BH-galaxy co-evolution a frontier of astronomy over the past two decades~(see reviews by \citealt{KH2013,Heckman2014}). These correlations themselves offer a new and more convenient approach to estimate the \mbh\ since the kinematic method used in local massive galaxies cannot be extended to more distant or dwarf galaxies.

\begin{figure*}
\includegraphics[scale=0.42]{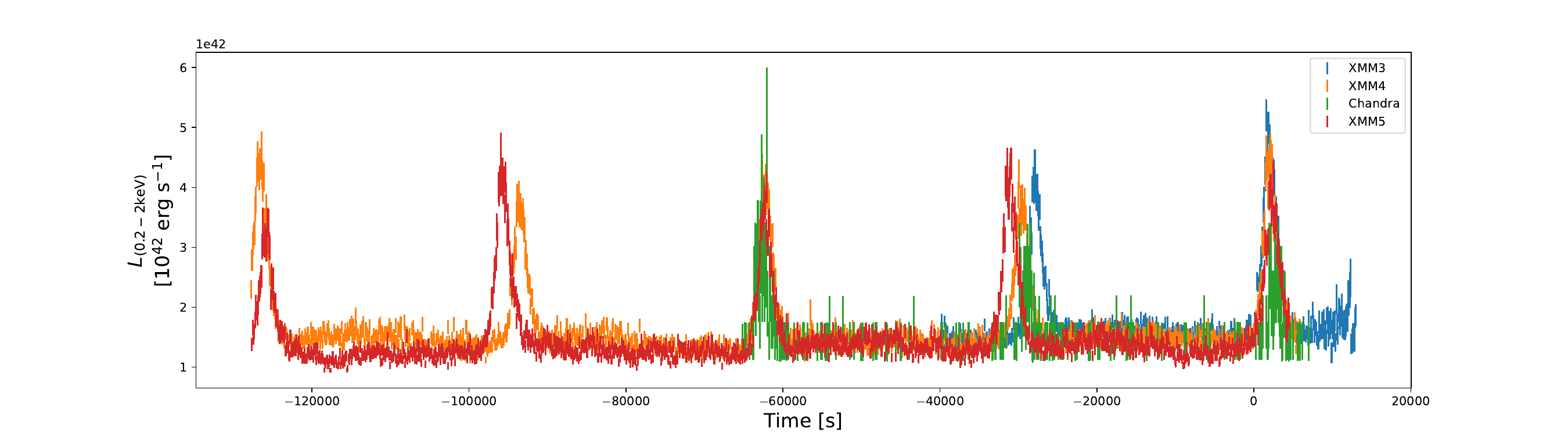}
\caption{\label{fig:lcs}  GSN 069  QPE light curves from 4 observations during Dec. 2018-May 2019 \citep{Miniutti2019,Miniutti2023b}, 
where the recurrence times show an alternating  long-short pattern $T_{\rm long, short}$, with both  $T_{\rm long}$ and $T_{\rm short}$ showing clear variations while the sum of two consecutive recurrence times 
$T_{\rm long} + T_{\rm  short} $ remains approximately a constant \citep{Zhou2024a}. } 
\end{figure*}

Although the $M_\bullet-\sigma_\star$ and the $M_\bullet-M_\star$ relations 
have been widely used to infer \mbh\ when the host bulge mass $M_\star$ or the stellar velocity dispersion $\sigma_{\star}$ is available, it is important to recognize their uncertainties, i.e. $0.3-0.4$ dex intrinsic scatter even when only considering elliptical galaxies and classical bulges~\citep{Gultekin2009,KH2013,McConnell2013}. 
Furthermore, these relations are primarily established based on SMBHs with $\mbh\gtrsim10^7$~\msun, leading to larger uncertainties for lower mass black holes, i.e. $\mbh\lesssim10^6$~\msun, either due to a larger intrinsic scatter or a possible systematic error.

For active galactic nuclei (AGNs), perhaps the most popular method of measuring \mbh\ is the virial mass,
which assumes that the broad-line region (BLR) is virialized and the motions of clouds are governed by the gravity of SMBH (see reviews by \citealt{Shen2013,Peterson2014}). There is a vague coefficient $f$ in the calculation whose average value is calibrated by the \mbh-$\sigma_{\star}$ relation. It means, in principle, that the virial mass estimator can not be more accurate than \mbh-$\sigma_{\star}$ relation. Moreover, $f$ is likely to vary from object to object, since it depends on the kinematics, geometry, inclination of the clouds, and even the types of the bulges~\citep{Ho2014}.  Recently, the evolutionary NIR interferometry GRAVITY, mounted on the Very Large Telescope Interferometer (VLTI), has opened up a new era of probing the BLR structure and has great potential to improve the precision of the \mbh\ measurement~\citep{GRAVITY2018,GRAVITY2024}, but it can only apply to very few $K$-band luminous AGNs. In addition, many other observational properties of AGNs, i.e. X-ray variability amplitude \citep{McHardy2006, Ponti2012, Pan2015} and optical variability time scale~\citep{Burke2021}, have been suggested to correlate with \mbh, though their underlying physics and uncertainties remain poorly understood.

In this paper, we propose a novel method to measure the \mbh\ with unprecedented precision using the quasi-periodic eruption (QPE) phenomenon. 
QPEs are intense repeating soft X-ray bursts with recurrence times about a few hours
to a few weeks from  galactic nuclei nearby. 
Starting from the first detection more than a decade ago \citep{Sun2013}, QPEs  from about ten different nearby galactic nuclei 
have been reported \citep{Miniutti2019,Giustini2020,Arcodia2021,Arcodia2022,Chakraborty2021,Evans2023,Guolo2024,Arcodia2024,nicholl2024,
Chakraborty:2025ntn, Hernandez-Garcia:2025ruv}.
The QPEs are detected in the soft X-ray band with similar peak luminosities ($10^{42}-10^{43}$ ergs s$^{-1}$),
thermal-like X-ray spectra with temperature $kT \simeq 100-250$ eV and 
the temperature  $50-80$ eV in the quiescent state.

\begin{figure}
\includegraphics[width=0.5\textwidth]{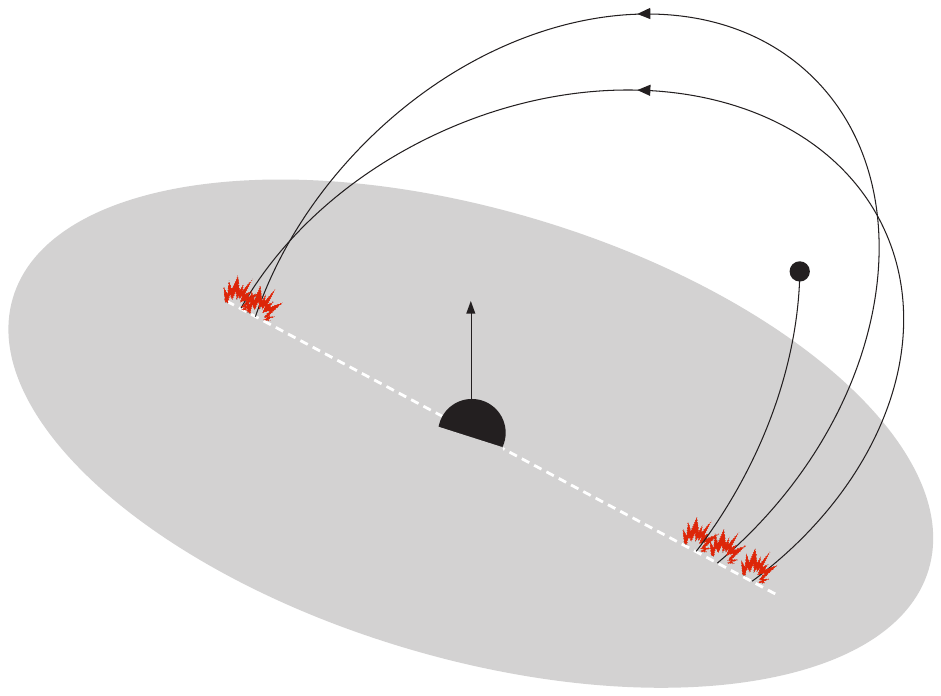}
\caption{\label{fig:cartoon} Schematic picture of the EMRI+disk model, where the EMRI collides the disk at a different location each time due to apsidal precession,
and the EMRI orbital plane precesses on an even longer Lense-Thirring precession timescale if the central SMBH is spinning. } 
\end{figure}

Though there have been some debates on the origin of QPEs, more and more analyses favor the extreme mass ratio inspiral (EMRI) + accretion disk model, where 
the QPEs are the result of collisions 
between a stellar mass object (SMO, a stellar mass black hole or a main sequence star) and an accretion disk
around a SMBH in galactic nuclei (see e.g., \citealt{Xian2021,Franchini2023,Tagawa:2023fpb,Arcodia2024,Guolo2024,Zhou2024a,Zhou2024b,Zhou2024c,Linial:2023xkx,Linial2023,Linial2023d, Chakraborty2024,Linial:2024mdz,Arcodia:2024taw,Yao:2024rtl,Giustini:2024dyy,Vurm:2024vfb,Pasham2024,Pasham:2024sox,Miniutti2025} for details).
Among all the observations,  two pieces of direct observational evidence for the EMRI+disk model have been recognized. Recently, QPEs in X-ray light curves of three tidal disruption events (TDEs) $\mathcal{O}(1)$ years after their ignitions have been directly detected \citep{nicholl2024,Bykov:2024ogf,Chakraborty:2025ntn}. Moreover, the QPE hosts and TDE hosts show strikingly  morphological similarities~\citep{Gilbert:2024ywq} and a preference for extending emission line regions indicative of recently faded AGNs~\citep{Wevers2024}.
The other observational evidence is about 
the alternating long-short pattern in the QPE recurrence times which has long been noticed in several QPE sources
\citep{Miniutti2019,Giustini2020,Arcodia2021,Arcodia2022}, including the most famous source GSN 069. 
A more intriguing feature in the QPE timing  was identified by \cite{Zhou2024a}: there are large variations in both the long and the short recurrence times of GSN 069 QPEs $T_{\rm long, short}(t)$, while $T_{\rm long}(t) +T_{\rm short}(t)$ remains nearly a constant (see Fig.~\ref{fig:lcs}). This observation strongly implies that 
$T_{\rm long} +T_{\rm short}$ is the fundamental period of underlying physical process 
that is sourcing the QPEs, and two flares with varying intervals are produced per fundamental period.
These two observations naturally fit in the EMRI+disk model (see Fig.~\ref{fig:cartoon}), while are hardly explained by other models in a natural way. Therefore, a unified model proposes that both TDEs and QPEs are the embers of AGNs, where AGNs increase both the TDE rate and the formation rate of low eccentricity EMRIs, and QPEs are preferentially found in recently faded AGNs, where TDEs often feed a misaligned accretion disk to the EMRI~\citep{Jiang2025}.

In the framework of EMRI+disk model, a range of astrophysical applications of QPEs have been investigated, 
including measuring SMBH parameters, probing structure of SMBH accretion disks, formation processes and formation rates of EMRIs \citep{Xian2021,Zhou2024a, Zhou2024b, Zhou2024c, Arcodia:2024efe,Kaur:2024ofj}
and implications for potential multi-messenger observations of EMRIs in the era of spaceborne gravitational wave astrophysics \citep{Kejriwal:2024bna}.
In this work, we will focus on measuring SMBH masses with QPE timing data, which we have briefly discussed in previous works \citep{Zhou2024a,Zhou2024b,Zhou2024c} (hereafter papers I-III). Using the QPE timing method, one can reconstruct the trajectory of a single SMO
around a SMBH, therefore accurately measure the orbital parameters and the SMBH mass.

The QPE timing method is a truly direct dynamic measurement achieved by tracing the motion of a single star orbiting around a SMBH, thereby circumventing many uncertainties related to both measurement and systematic errors involved in traditional methods. This method is in the same spirit of measuring the Sagittarius $\rm A^{\star}$  SMBH mass by monitoring  orbits of individual stars ~\citep{Schodel2002,Ghez2005,Gillessen2009,GRAVITY2018b}.
As QPEs are preferred to be found in dwarf galaxies hosting  low-mass black holes ~\citep{Wevers2022}, this technique is particularly valuable as all previous methods are much less accurate in the low mass range~\citep{Greene2020}. For example, the mass \mbh\ of the prototype and most well-studied intermediate-mass black hole (IMBH) in the nearby AGN NGC~4395 has a huge mass uncertainty between $4\times10^{5}$~\msun\ and $1\times10^4$~\msun\ reported by various works~\citep{Peterson2005,denBrok2015,Woo2019}. In this regard, the QPE timing method we proposed will also aid in the detection of IMBHs with robust mass measurement, particularly in inactive galaxies throughout the universe. 

Existing light curve data of QPEs have been obtained using XMM-Newton \citep{Miniutti2019,Miniutti2023,Miniutti2023b, Arcodia2021, Arcodia:2024taw}, Chandra \citep{Miniutti2023b} and NICER \citep{Arcodia2021, Chakraborty2024,nicholl2024}. Among these instruments, XMM-Newton provides the highest quality data but its observation epochs typically span only about one day. The precision of mass measurement is expected to improve substantially with extended-duration monitoring.

The Chasing All Transients Constellation Hunters (CATCH) \citep{catch}, proposed by the Institute of High Energy Physics at the Chinese Academy of Sciences, comprises a constellation of $\sim100$ satellites aiming to conduct simultaneous follow-up observations for diverse transients in the X-ray band. On June 22, 2024, the first CATCH pathfinder \citep{catch-1} was launched alongside the Space-based multiband astronomical Variable Objects Monitor (SVOM) mission \citep{svom}. The complete constellation is projected to be deployed in the early 2030s. CATCH uses the WALKER constellation design with 3 orbital planes, each of which consists of a few tens of satellites. This configuration enables uninterrupted (long-term) monitoring mode~\citep{catch}, effectively allowing continuous observation of a source. A small number of satellites can be fully reserved for QPE observations.\footnote{Private communication with Lian Tao, PI of CATCH.} In this paper, we forecast the potential contribution of CATCH to the $M_\bullet$ measurement and compare its capabilities with those of current instruments.

This paper is organized as follows. In Section~\ref{sec:princ}, we illustrate the underlying principles of the QPE timing method.
In Section~\ref{sec:model}, we introduce the details of the Bayesian framework for QPE timing analysis.
In Section~\ref{sec:Strategy}, we simulate four strategies to demonstrate the capability of measuring $M_\bullet$ of QPE timing method and two more strategies to demonstrate the unique potential of CATCH to measure $a$.
Summary and discussions are given in Section~\ref{sec:conclusion}.
Throughout this paper, we will use geometric units with convention $G=c=1$.

\section{Basic principles of the QPE timing method}\label{sec:princ}
The physical picture of constraining the SMBH mass from QPE timing is rather simple. As the SMO orbiting around the SMBH, it crosses the accretion disk twice and produces two flares per orbit.
Therefore, there is an alternating long-short pattern in the QPE recurrence times, arising from the orbital eccentricity and 
different light path length from the two collision locations to the observer. The orbital period can be identified as $T_{\rm obt}=T_{\rm long}+T_{\rm short}$ to a good precision, where $T_{\rm long}$ and $T_{\rm short}$ are the adjacent QPE recurrence times. 
Different from in Newtonian gravity, the EMRI orbit in the curved spacetime does not close itself due to the apsidal precession.
As a result, both $T_{\rm long}$ and $T_{\rm short}$ are time dependent and vary on the timescale of the apsidal precession period $T_{\rm aps}$.
Both periods $T_{\rm obt}$ and $T_{\rm aps}$ can be constrained with a reasonable amount of QPE timing data, and are related to the SMBH mass $M_\bullet$ and the semi-major axis $A$ of the EMRI orbit via the Kepler's third law
\be \label{eq:kepler}
\begin{aligned}
    T_{\rm obt}  = 2 \pi \left(\frac{A}{M_\bullet}\right)^{3/2} M_\bullet\ , \\ 
\end{aligned}
\ee 
and the ratio
\be \label{eq:aps_obt}
\frac{T_{\rm aps}}{T_{\rm obt}} =  \frac{p}{3M_\bullet}  \approx \frac{A}{3M_\bullet}\ ,
\ee 
where $p$ is the semi-latus rectum relating to the semi-major axis $A$ and the orbital eccentricity $e$ as $A=p/(1-e^2)$,  the approximation sign in the equation above is accurate for low-eccentricity  orbits.
With the two observables $T_{\rm obt}$ and $T_{\rm aps}$, one can naturally constrain the SMBH mass with $M_\bullet \propto T_{\rm obt}^{5/2} T_{\rm aps}^{-3/2}$. 

Note that the two periods measured in the observer frame are different from their intrinsic values by a factor of $(1+z)$ due to the cosmology redshift $z$,
therefore the SMBH mass $M_\bullet$  derived above is in fact a redshifted mass, which is different from the intrinsic mass by a factor of $(1+z)$. 
The differences are small for currently detected QPE sources at low redshifts $z\sim \mathcal{O}(0.01)$, but could be important for sources at high redshifts.

In a similar way, one can constrain the Lense-Thirring precession period $T_{\rm LT}$ of the EMRI orbit from the QPE timing data 
if the QPE source has been monitored for a sufficiently long time $\gtrsim T_{\rm LT}$. Consequently, one
can infer the dimensionless spin $a$ of the SMBH via the relation
\be \label{eq:LT_obt}
\frac{T_{\rm LT}}{T_{\rm obt}} = \frac{1}{2a}\left(\frac{p}{M_\bullet}\right)^{3/2}\ .
\ee 
In practice, 
we choose to conduct the QPE timing analysis in the Bayesian framework, the basic ingredients of which have been shown in previous works (papers I-III).
In this work, we aim to generalize the previous analyses and construct a QPE timing method which is capable of measuring orbital parameters and SMBH masses $M_\bullet$ (and spins $a$) in a comprehensive Bayesian framework.

To summarize, both the SMBH mass and spin can be obtained from the three periods $T_{\rm obt}, T_{\rm aps}, T_{\rm LT}$ via Eqs.~(\ref{eq:kepler}-\ref{eq:LT_obt}), 
which are intrinsic properties of Kerr geodesics and are independent of the existence of an accretion disk or not. 
In the following section, we will explain the details of how to measure these three periods along with disk motion parameters from QPE observations of an EMRI+disk system, where the disk could either be an equatorial disk or a misaligned and precessing  disk.

\section{Bayesian framework for QPE timing analysis}\label{sec:model}

As discussed in previous studies, one can predict the SMO-disk collision times and the resulting QPE light curve with a model of SMO motion and an accretion disk model.
But the full light prediction is subject to large uncertainties in the disk model, the nature of the SMO and the radiation mechanism. 
We therefore choose to constrain the EMRI kinematics and the QPE emission separately for mitigating the impact of these uncertainties.
We first fit each QPE with a simple light curve model and obtain the starting time of each flare $t_0\pm \sigma(t_0)$ (see papers I-III for details), which is identified as the time of the SMO crossing the disk and is used for constraining the SMO orbital parameters. With data $d=\{t_0^{(k)}\pm \sigma^{(k)}(t_0)\}$ ($k$ is the flare index) and a  QPE flare timing model, we can constrain model parameters $\mathbf{\Theta}$ in the Bayesian inference framework. According to the Bayes theorem, the posterior of model parameters is written as 
\be \mathcal P(\mathbf{\Theta}, \mathcal{H}|d) =\frac{ \mathcal{L}(d|\mathbf{\Theta},\mathcal{H}) \pi(\mathbf{\Theta},\mathcal{H})}{\mathcal{Z}(d)}\ , \ee
where $\mathcal{L}(d|\mathbf{\Theta},\mathcal{H})$ is the likelihood of detecting data $d$ under hypothesis $\mathcal{H}$
with model parameters $\mathbf{\Theta}$, $\pi(\mathbf{\Theta},\mathcal{H})$ is the assumed prior for model parameters $\mathbf{\Theta}$ 
in $\mathcal{H}$,
and the normalization factor $\mathcal{Z}(d)$ is the evidence of hypothesis $\mathcal{H}$ with data $d$.

To quantify the support for one hypothesis $\mathcal{H}_1$ over another $\mathcal{H}_0$  by  data $d$,
we can calculate the evidence ratio of two hypotheses, i.e., the Bayes factor,  
\be 
\mathcal{B}_0^1 = \frac{ \mathcal{Z}_1(d)}{\mathcal{Z}_0(d)}\ .
\ee
The larger $\mathcal{B}_0^1$ represents stronger support for hypothesis  $\mathcal{H}_1$ over $\mathcal{H}_0$.
In Jeffreys' scale, $\log\mathcal{B}_0^1\in (1.2, 2.3), (2.3, 3.5), (3.5, 4.6), (4.6, \infty)$ are the criteria of substantial, strong, very strong,
and decisive strength of evidence, respectively.

In the following subsections, we will explain the two major components of the QPE timing model (EMRI trajectories and disk motion), define the likelihood for the Bayesian framework of the QPE timing method for measuring SMBH masses. 

\subsection{Forced EMRI trajectories}
\label{sec:forced_EMRI}

As shown in paper III,  the SMO orbital energy dissipation  as crossing the accretion disk is small but can be measured 
in some QPE sources via long term monitoring. In this subsection, we briefly review the steps of computing 
forced EMRI trajectories in the Kerr spacetime. We start from time-like geodesics in the Kerr spacetime, analytic solutions to which
have been derived by \cite{Fujita:2009bp} and \cite{vandeMeent:2019cam}
as 
\be 
\begin{aligned}
    r(\lambda) &= r(q_r(\lambda); E, L, C)\ , &q_r(\lambda) &= \Upsilon_r\lambda+q_{r,\rm ini}\ , \\ 
    z(\lambda) &= z(q_z(\lambda); E, L, C) \ , &q_z(\lambda) &= \Upsilon_z\lambda+q_{z,\rm ini}\ ,\\
    t(\lambda) &= t(q_{t,r,z}(\lambda); E, L, C)\ , &q_t(\lambda) &= \Upsilon_t\lambda+q_{t,\rm ini} \ , \\
    \phi(\lambda) &= \phi(q_{\phi,r,z}(\lambda); E, L, C)\ , &q_\phi(\lambda) &= \Upsilon_\phi\lambda+q_{\phi,\rm ini}\ ,\\
\end{aligned}
\ee 
where $\lambda$ is the Mino time; $\{E, L, C\}$ are the energy, angular momentum and the Carter constant, respectively;
$\{\Upsilon_r,\Upsilon_z,\Upsilon_t,\Upsilon_\phi\}$ are the Mino time frequencies in the $r, z(=\cos\theta), t, \phi$ direction, respectively; $\{q_{r,\rm ini},q_{z,\rm ini},q_{t,\rm ini},q_{\phi,\rm ini}\}$ are the initial phases.
The conversion relation between the constants $\{E, L, C\}$ and the orbital parameters $\{p, e, \cos\theta_{\rm min}\}$ can be found in \cite{Schmidt2002}, where $\theta_{\rm min}$ is the minimum polar angle that the geodesic can reach.
A geodesic is uniquely determined by a set of orbital parameters $\{p, e, \cos\theta_{\rm min}\}$ and initial phases 
$\{q_{r,\rm ini},q_{z,\rm ini},q_{\phi,\rm ini}\}$ at initial time $t_{\rm ini}$, where the initial phase $q_{t,\rm ini}$
is fixed by $t(q_{t,\rm ini},q_{r,\rm ini},q_{z,\rm ini})=t_{\rm ini}$.

Assuming the orbital energy dissipation as the SMO crosses the accretion disk is a small perturbation, 
then the SMO trajectories can be formulated as perturbed geodesics
with varying constants  $\{E(t), L(t), C(t)\}$. 
In the language of perturbation techniques, the forced EMRI trajectories can be computed using the method of osculating orbits, i.e.,
 in the adiabatic approximation, the equations of motion can be  written as 
\be 
\begin{aligned}
    r(\lambda) &= r(q_r(\lambda); E, L, C)\ , &\dv{q_r}{\lambda} &= \Upsilon_r(E,L,C)\ , \\ 
    z(\lambda) &= z(q_z(\lambda); E, L, C) \ , &\dv{q_z}{\lambda} &= \Upsilon_z(E,L,C)\ ,\\
    t(\lambda) &= t(q_{t,r,z}(\lambda); E, L, C)\ , &\dv{q_t}{\lambda} &= \Upsilon_z(E,L,C) \ , \\
    \phi(\lambda) &= \phi(q_{\phi,r,z}(\lambda); E, L, C)\ , &\dv{q_\phi}{\lambda} &= \Upsilon_\phi(E,L,C) \ ,\\
\end{aligned}
\ee 
with time-dependent $\{E, L , C\}$ and  thereby time-dependent frequencies $\Upsilon_{r,z,t,\phi}$.

The time dependence of $\{E, L , C\}$ can be derived from the change of orbital parameters of $\{T_{\rm obt}(p,e), e, \cos\theta_{\rm min}\}$ as the SMO crosses the disk. As shown in several recent studies of star-disk collisions \cite[e.g.,][]{Linial2023,Wang2023},  
the relative changes in orbital parameters are similar in magnitudes with $\delta e/e \sim \delta T_{\rm obt}/T_{\rm obt}\sim \delta \theta_{\rm min}/\theta_{\rm min}$. As we have shown in previous studies, $T_{\rm obt}$ is the best constrained orbital parameter
while $e$ and $\cos\theta_{\rm min}$ are less constrained with $\mathcal{O}(1)$ fractional uncertainties.
As a result, the small fractional change in the orbital period $\delta T_{\rm obt}$ is detectable, 
while $\delta e$ and $\delta \theta_{\rm min}$ are undetectable for the QPE sources currently available.
We therefore can safely take $\dot e = \dot \theta_{\rm min}=0$ in calculating the EMRI trajectories.
As for the orbital period decay rate $\dot T_{\rm obt}(t)$,  we  model it as 
\be \label{eq:Tdot}
\dot T_{\rm obt}(t) = \dot T_{\rm obt, max} \sin\iota_{\rm sd}(t)\ ,
\ee
where $\iota_{\rm sd}$ the angle between the SMO orbital plane and the disk plane.
This function form is motivated by the energy loss for a star crossing an accretion disk \citep{Linial:2023xkx,Linial2023}.
For a precessing misaligned disk, $\iota_{\rm sd}$ is modulated by the disk precession and the SMO orbital precession, therefore the decay rate $\dot T_{\rm obt}$ is non-uniform. In the case of an equatorial accretion disk, the orbital period decay rate simplifies as a constant. 

\subsection{Disk motion: precession and alignment}\label{subsec:disk motion}

In general, a TDE star that is scattered into the tidal radius of a SMBH is from a random direction.
As a result, the initial orientation of the accretion disk formed in a TDE is also random. 
The possible signature of the disk precession on the QPE timing has been discussed in several previous studies \citep[e.g.,][]{Franchini2023, Chakraborty2024,Arcodia:2024taw,Miniutti2025}.
As a minimal assumption of the disk precession, we model it as a rigid body like precession with a constant precession rate. The normal vector of the disk plane is written as 
\be 
\vec n_{\rm disk}=(\sin\beta\cos\alpha, \sin\beta\sin\alpha, \cos\beta) \ ,
\ee 
where $\alpha\in (0, 2\pi)$ is the azimuth angle and $\beta\in (0, \pi/2)$ is the angle between the disk plane and the equatorial plane. 
With the constant precession rate assumption, the azimuth angle then evolves as 
\be \label{eq:disk_precession}
    \alpha(t) = \alpha_{\rm ini}+\frac{2\pi}{\tau_{\rm p}}(t-t_{\rm ini})\ , \\
\ee 
where $\tau_{\rm p}$ is the disk precession period, and $\alpha_{\rm ini}$ is  the initial value of the azimuth angle  at $t_{\rm ini}$. 

In the long run, the initially misaligned disk is expected to approach the equator plane gradually.
One can in principle constrain the evolution history of the inclination angle  $\beta(t)$  in a non-parametric approach
if sufficiently dense observations of QPEs are available. However, QPE sources are usually sparsely monitored due to  limited 
X-ray observation resources.  In practice, we choose to bridge the spare observations with   
a simple function
\be \label{eq:align}
\sin\beta (t) = \sin\beta_{\rm ini} \exp\left\{-\frac{t-t_{\rm ini}}{\tau_{\rm a}}\right\}\ ,
\ee 
parameterizing the alignment process with the alignment timescale parameter $\tau_{\rm a}$, where $\beta_{\rm ini}$ in the initial value of the disk inclination angle. This function form is motivated by the eigen-mode analysis of disk alignment process \citep[e.g.,][]{Scheuer1996,Zanazzi:2019ujv}.

 \subsection{QPE timing model}
In the previous two subsections, we have summarized the key points in modeling the EMRI motion and the disk motion,
based on which one can construct a full QPE timing model. 
But not all the ingredients are necessary for modeling each QPE source,
e.g.,  no evidence for disk precession in GSN 069 QPEs is found \citep{Zhou2024c}.
Therefore we consider the following two hypotheses with slightly different assumptions about the EMRI motion and the disk motion, 
and the data favored hypothesis will be selected by Bayesian analyses:

\begin{enumerate}
    \item Vanilla hypothesis ($\mathcal{H}_0$): The SMO moves around the SMBH losing orbital energy as crossing the equatorial accretion disk. The EMRI+disk system 
    can be specified by $9$ parameters: the intrinsic orbital parameters $(p, e, \theta_{\rm min})$, the initial phases $(q_{r,{\rm ini}}, q_{z,{\rm ini}}, q_{\phi,{\rm ini}})$, the mass of the SMBH $M_\bullet$ or equivalently the orbital period $T_{\rm obt}$ [Eq.~(\ref{eq:kepler})], the dimensionless spin of the SMBH $a$ and the orbital period decay rate $\dot T_{\rm obt}$.
    \item Disk precession and alignment hypothesis ($\mathcal{H}_1$): Different from the vanilla hypothesis, we consider a misaligned disk with initial
    orientation angles $(\alpha_{\rm ini}, \beta_{\rm ini}$) and precessing around the polar direction with a constant period $\tau_{\rm p}$.
    In this hypothesis, the angle between the SMO orbital plane and the accretion disk $\iota_{\rm sd}$ is time dependent, so does the orbital period decay rate [Eq.~(\ref{eq:Tdot})].
    In addition to disk precession, we also consider a possible disk alignment process which is parameterized with an alignment timescale $\tau_{\rm a}$ [Eq.~(\ref{eq:align})].
    Therefore, 4 additional parameters $\{\alpha_{\rm ini}, \beta_{\rm ini},\tau_{\rm p}, \tau_{\rm a}\}$ are introduced for describing the disk motion in this hypothesis (and the orbital period decay rate
    parameter $\dot T_{\rm obt}$ in $\mathcal{H}_0$ is replaced with $\dot T_{\rm obt, max}$).
\end{enumerate}

From a SMO trajectory and disk motion, one can calculate the disk crossing times $t_{\rm crs}$,
which we identify as the flare starting times. Specifically, we choose the disk crossing time as  when the SMO crosses the upper disk surface or the lower disk surface depending on the observer direction, i.e. $r_{\rm crs}(\vec n_{\rm crs}\cdot \vec n_{\rm disk})=H \ {\rm sign}(\vec n_{\rm obs}\cdot \vec n_{\rm disk})$. 
Without loss of generality, we fix the observer in the $x-z$ plane, i.e., the unit direction vector pointing to the observer is $\vec n_{\rm obs} = (\sin\theta_{\rm obs}, 0, \cos\theta_{\rm obs})$.
The propagation times of different flares at different collision locations $r_{\rm crs} \vec n_{\rm crs}$
to the observer will also be different. Taking the light propagation delays into account,
we  find the flare starting time in the observer frame as  
\be\label{eq:tobs_theo}
t_{\rm obs} = t_{\rm crs} + \delta t_{\rm geom} + \delta t_{\rm shap} \ ,
\ee 
where 
\be\label{eq:tobs}
\begin{aligned}
    \delta t_{\rm geom} &= -r_{\rm crs} \vec n_{\rm obs}\cdot \vec n_{\rm crs}\ , \\ 
    \delta t_{\rm shap} &=-2M_\bullet\log \left[r_{\rm crs} (1+ \vec n_{\rm obs}\cdot \vec n_{\rm crs})\right]\ ,
\end{aligned}
\ee 
are corrections caused by different light path lengths and different Shapiro delays \citep{Shapiro1964}, respectively.
To summarize, our QPE timing model is written as $t_{\rm obs}^{(k)}(\mathbf{\Theta},\mathcal{H})$ in a short notation, 
where $k$ is the index of observed flares, $\mathcal{H}$ is the hypothesis adopted and $\mathbf{\Theta}$ is the associated model parameters.

\subsection{Bayesian framework}

As discussed in previous analyses, it is possible that there are some physical processes that affect the QPE timing but are not included in our QPE timing model.
Assuming the unmodeled advances or delays in the  QPE timing follows a Gaussian distribution with variance $\sigma_{\rm sys}^2$,
the likelihood of seeing data $d=\{t_0^{(k)}\}$ under hypothesis $\mathcal{H}$ with model parameters $\mathbf{\Theta}$ is written as 
(see paper II for a short derivation)
\be \label{eq:likeli}
\mathcal{L}_{\rm timing}(d|{\mathbf{\Theta}}, \mathcal{H})=\prod_{k}\frac{1}{\sqrt{2\pi (\tilde\sigma(t_0^{(k)}))^2}}
 \exp\left\{-\frac{(t_{\rm obs}^{(k)}- t_0^{(k)})^2}{2(\tilde\sigma(t_0^{(k)}))^2} \right\} \ ,
\ee 
where $(\tilde\sigma(t_0^{(k)}))^2=(\sigma(t_0^{(k)}))^2+\sigma_{\rm sys}^2$ is the uncertainty contributed by both measurement errorbars and unmodeled uncertainties.
Note that $\mathbf{\Theta}$ includes both the physical parameters introduced in the previous subsection and the systematic uncertainty parameter $\sigma_{\rm sys}$.
Similar inference method has been widely used in the context of hierarchical test of General Relativity with gravitational waves  \citep{Isi2019}.


For each QPE source, we apply both  hypotheses $\mathcal{H}_{0,1}$ to the timing data of observed flares $t_0^{(k)}$,  
and perform model parameter inferences using the \texttt{nessai} \citep{nessai} algorithm within \texttt{Bilby} \citep{Ashton2019} with the default settings, except more live points \texttt{nlive}=2000 for better performance in the parameter inference.
As we will see later, the data favored hypothesis will be featured with higher evidence obtained from the Bayesian inference, and the 
hypothesis preference is quantified by the Bayes factor.

As a clean measurement of the SMBH mass from QPE timing data alone, 
we do not include any prior information about the SMBH mass informed by external measurements, e.g., commonly used  $M_\bullet-\sigma_\star$ relations \citep{Tremaine2002, Gultekin2009}.

\section{Applications of the QPE timing method}\label{sec:Strategy}

\subsection{GSN 069}

\begin{table}
    \centering
    \resizebox{0.8\columnwidth}{!}{%
    \begin{tabular}{l|cc}
       $\mathbf{\Theta}$ & $\pi(\mathbf{\Theta}, \mathcal{H}_1)$ & $\pi(\mathbf{\Theta},\mathcal{H}_0)$ \\
        \hline
       $p\ [M_\bullet]$ & $ \mathcal{U}[50, 500]$ & \\ 
       $e$ & $\mathcal{U}[0, 0.9]$ & \\
       $\cos(\theta_{\rm min})$ & $\mathcal{U}[0, 1]$ & \\
       $q_{r,{\rm ini}}$ & $\mathcal{U}[0, 2\pi]$ & \\
       $q_{z,{\rm ini}}$ & $\mathcal{U}[0, 2\pi]$ & \\
       $q_{\phi,{\rm ini}}$ & $\mathcal{U}[0, 2\pi]$ & \\
       $T_{\rm obt}\ [{\rm ks}]$ & $\mathcal{U}[60, 70]$ & \\
       $a$ & $\mathcal{U}[0, 1]$ & \\
       $\theta_{\rm obs}$ & $\mathcal{U}[0, \pi]$ & $\mathcal{U}[0, \pi/2]$ \\
       $\alpha_{\rm ini}$ & $\mathcal{U}[0, 2\pi]$ &  None \\
       $\tau_{\rm p} \ [{\rm days}]$ & $\mathcal{U}[1, 2000]$ & None \\
       $\beta_{\rm ini}$ & $\mathcal{U}[0, \pi/2]$ & None \\
       $\tau_{\rm a} \ [{\rm days}]$ & $\mathcal{U}[1, 2000]$ & None \\
       $\dot{T}_{\rm obt} \ [\times10^{-5}]$ & None & $\mathcal{U}[-10, 0]$ \\
       $\dot{T}_{\rm obt,max} \ [\times10^{-5}]$ & $\mathcal{U}[-10, 0]$ & None \\
       $\sigma_{\rm sys} \ [{\rm ks}]$ & $\mathcal{U}[0, 2]$ & 
    \end{tabular} }
    \caption{Priors used for the orbital parameter inference of GSN 069 EMRI. Entries left blank in the $\pi(\mathbf{\Theta}, \mathcal{H}_0)$ column indicate that the same prior values from the $\pi(\mathbf{\Theta}, \mathcal{H}_1)$ column are imposed.
    } 
    \label{tab:gsn_prior}
\end{table}

\begin{figure*}
\includegraphics[scale=0.4]{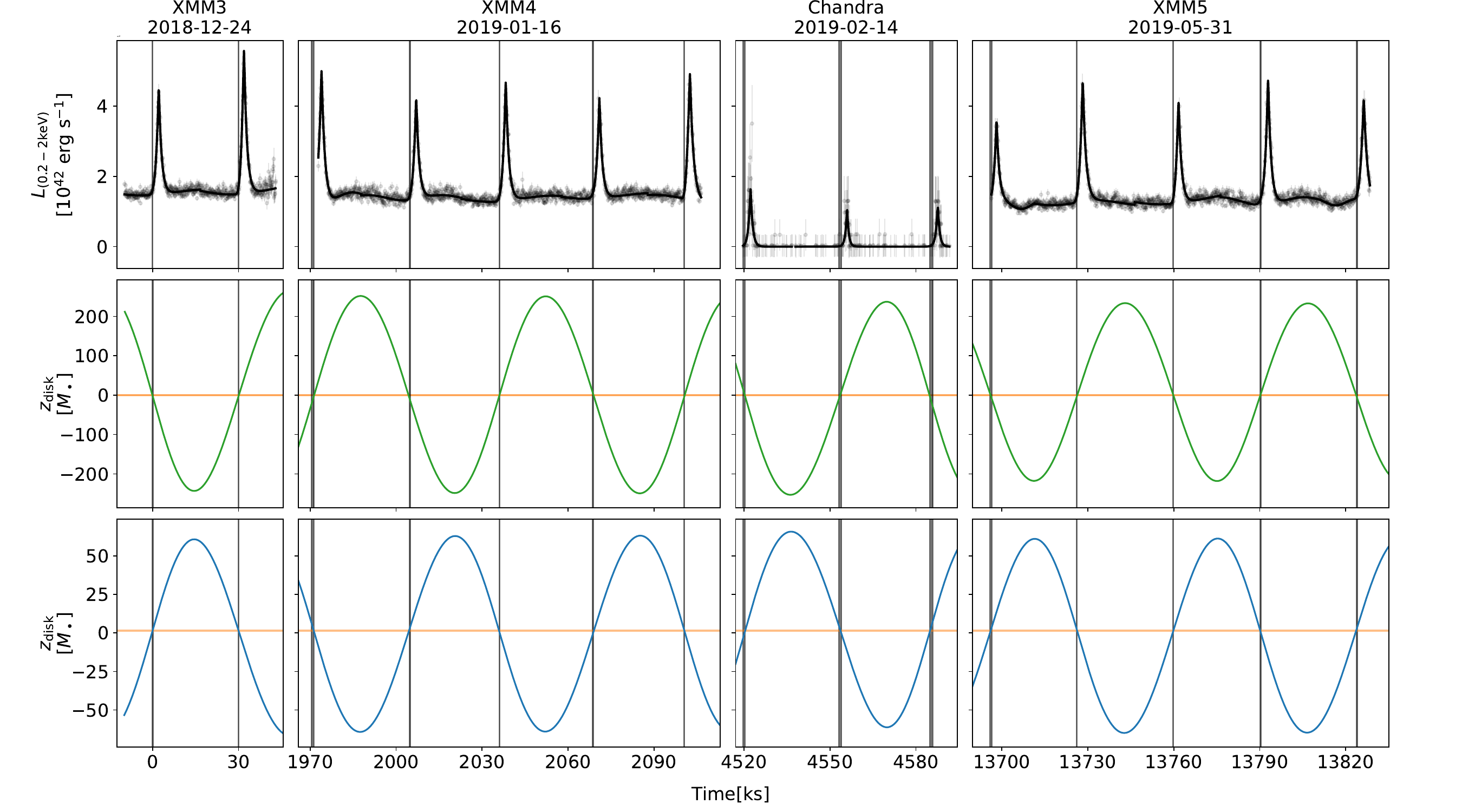}
\caption{\label{fig:lo_GSN} Top panel: light curve data of GSN 069 along with the best-fit EMRI trajectories, where the vertical bands are the inferred starting times $t_0^{(k)}\pm \sigma(t_0^{(k)})$ of the QPEs. Middle panel: distance to the disk midplane $z_{\rm disk}(t)$ of the best-fit orbits for the disk precession and alignment hypothesis ($\mathcal{H}_1$), where the orange horizontal lines denote the disk surface $z=H$ and the verticals bands are the inferred starting times $t_0^{(k)}\pm \tilde\sigma(t_0^{(k)})$, with
$\tilde\sigma(t_0^{(k)})=\sqrt{(\sigma(t_0^{(k)}))^2+\sigma_{\rm sys}^2}$. Bottom panel: same to the midle panel but for the vanilla hypothesis ($\mathcal{H}_0$).
}
\end{figure*}
We first apply the comprehensive Bayesian framework to the well-studied QPE source, GSN 069, which has been extensively analyzed in paper I-III. In Table~\ref{tab:gsn_prior}, we present the priors of model parameters used for orbital analyses in $\mathcal{H}_0$, $\mathcal{H}_1$. 
In Fig.~\ref{fig:lo_GSN}, the best-fit EMRI
trajectories of $\mathcal{H}_0$ and $\mathcal{H}_1$ are displayed along with
the QPE light curves.
The posterior corner plots of all model parameters are shown in Fig.~\ref{fig:gsn_h0} and Fig.~\ref{fig:gsn_h2}, respectively.
The orbital parameters are constrained as 
\be 
\begin{aligned}
    p &= 304^{+56}_{-134}  \ M_\bullet\ , \\ 
    e &= 0.04^{+0.03}_{-0.02}  \ ,\quad (\mathcal{H}_0) \\
    T_{\rm obt} &= 64.73^{+0.02}_{-0.03} \ {\rm ks}\ , \\
    \dot T_{\rm obt} &=-6.5^{+0.2}_{-0.2}\times10^{-5}\ , \\ 
    T_{\rm aps} &= 76^{+14}_{-34} \ {\rm days} \ ,
\end{aligned}
\ee 
and 
\be \label{eq:H1_constraint}
\begin{aligned}
    p &= 300^{+193}_{-48} \ M_\bullet\ , \\ 
    e &= 0.04^{+0.02}_{-0.03} \ , \\
    T_{\rm obt} &= 64.71^{+0.03}_{-0.05} \ {\rm ks}\ , \quad (\mathcal{H}_1)\\ 
    \dot T_{\rm obt,max} &=-7.3^{+0.9}_{-2.2}\times10^{-5}\ , \\
     T_{\rm aps} &= 75^{+48}_{-12} \ {\rm days} \ ,\\
    \tau_{\rm p} &= 1288^{+664}_{-736} \ {\rm days} \ , \\
    \tau_{\rm a} &= 1144^{+803}_{-810} \ {\rm days} \ , 
\end{aligned}
\ee 
at 2-$\sigma$ confidence level, where the apsidal precession period is obtained from Eq.~(\ref{eq:aps_obt}). 
The log Bayes factor between the two hypotheses is found to be
\be\label{eq:logB}
    \log\mathcal{B}^1_0 =-1.9\pm0.2 \ .
\ee 
In Jeffreys’ scale,  disk precession in GSN 069 is substantially disfavored.
\footnote{ \citet{Miniutti2025} claimed that the super-orbital modulation in the GSN 069 data is not likely from EMRI apsidal precession, but is from disk precession. Probably the most reliable way to test this claim  is a simultaneous measurement of the EMRI apsidal precession period $T_{\rm aps}$ and the disk precession period $\tau_{\rm p}$ from the data assuming EMRI+a precessing disk ($\mathcal{H}_1$).  However, the measurement result  $\tau_{\rm p} \gg T_{\rm aps}\approx 76$ day [Eq.~(\ref{eq:H1_constraint})] supports the opposite: the disk precession is slow. NOTE that the $\tau_{\rm p}$ and $T_{\rm aps}$ constraints are measured from the GSN 069 data in the framework of EMRI+disk model, instead of an assumption. In Appendix~\ref{app_b}, we perform detailed O-C analyses for comparison and discuss possible sources of discrepancy in
interpreting the super-orbital modulation in the GSN 069 data.}

The Bayes factor itself may not be conclusive in excluding disk precession, since it depends on the parameter priors used. 
As a cross check, one can directly measure the disk precession period $\tau_{\rm p}$ from the QPE data. 
Consistent with the indication of the Bayes factor above, the disk precession period $\tau_{\rm p}$ is found to be even longer than the observation span [see Eq.~(\ref{eq:H1_constraint})], i.e.,
the disk precession is slow and has introduced little modulation on the QPE timing within the observation span. In addition, the two best-fit trajectories in hypotheses $\mathcal{H}_0$ and $\mathcal{H}_1$ are quit similar, which is also consistent with slow disk precession in GSN 069.
Longer monitoring is needed for accurately measuring such slow disk precession, and we will investigate disk precession measurement accuracy in more details in the following subsection.

The SMBH mass constraints in the data favored hypothesis $\mathcal{H}_0$ is obtained from Eqs.~(\ref{eq:kepler},\ref{eq:aps_obt}) as  
\be \label{eq:GSN069_mass}
\begin{aligned}
    \log_{10}(M_\bullet/M_\odot) &= 5.6^{+0.4}_{-0.1}\ , \qquad (\mathcal{H}_0)
\end{aligned}
\ee 
at 2-$\sigma$ confidence level. 
This SMBH mass constraint is independent of and of significantly lower uncertainty than the constraint inferred from the $M_\bullet-\sigma_\star$ relation , $\mathrm{log}_{10}(M_{\bullet}/M_{\odot}) = 6.0\pm1.0$ (at 2-$\sigma$ confidence level) \citep{Wevers2022}. 

In the data less favored hypothesis $\mathcal{H}_1$, the apsidal precession and SMBH mass constraints are found to be
\be 
\begin{aligned}
    \log_{10}(M_\bullet/M_\odot) &= 5.6^{+0.1}_{-0.3}\ , \qquad (\mathcal{H}_1)
\end{aligned}
\ee 
which is consistent with Eq.~(\ref{eq:GSN069_mass}). This comparison also shows that the slow disk precession  in GSN 069 
has introduced weak modulation to the QPE timing. As a result, 
two different hypotheses yield consistent constraints of the SMBH mass.

GSN 069 represents a good example of dynamical measurement of SMBH masses with the QPE timing method.
In addition to intrinsic EMRI orbital period and apsidal precession period, constraints of disk motion parameters are also obtained 
as side products.

\subsection{Optimal observation strategy: SMBH mass measurement}

In this subsection, we simulate different strategies of observing QPEs and evaluate their performances in measuring SMBH masses. 
As an example, we consider a mock QPE source that is similar to GSN 069 and eRO-QPE2, with source parameters $M_\bullet=10^6 M_\odot, \ a=0.9, \ p= 100 M_\bullet, \ e=0.04$. 
The three dynamical timescales are therefore $T_{\rm obt}=31.5$ ks, $T_{\rm aps}=12.1$ days and $T_{\rm LT}=202.5$ days.
For convenience, we assume an equatorial disk and the injection values of full QPE timing model parameters are listed in Table~\ref{tab:stra_pri}.
More general cases of EMRI+ a precessing disk will be explored in a following subsection.

Using the osculating trajectory method described above, true flare starting times in the observer frame $t_{\rm obs, true}^{(k)}$ are calculated directly following Eq.~(\ref{eq:tobs_theo}).
The observed flare starting times $t_{0}^{(k)}$ will be slightly different due to measurement noises, 
\be
\begin{aligned}
    t_{0}^{(k)} &= t_{\rm obs, true}^{(k)} + \delta t_{0}^{(k)}\ , 
\end{aligned}
\ee
where $\delta t_{0}^{(k)}$ is randomly drawn from a Gaussian distribution with a mean value $0$ and a standard deviation $\sigma(t_{0}^{(k)})$. In this work, we fix the measurement uncertainty as  $\sigma(t_{0}^{(k)})=100$ s. 
\footnote{From XMM-Newton observations of eRO-QPE2, flare starting time uncertainties are found in the range of $\sim (30, 80)$ s \citep{Zhou2024b}. The current design of effective collecting area 
of a single CATCH satellite is $140$ cm$^2$ in the keV range, which is smaller than that of XMM-Newton by a factor of $\sim 8$. 
A simple scaling analysis shows that $\sim (80, 220)$ s or $\sim (60, 160)$ s uncertainties are expected if one or two CATCH satellites are used for eRO-QPE2 observations. We use $100$ s as a fiducial value of flare starting time uncertainties in this work and it is straightforward to obtain model parameter constraints by simple scaling if a different resolution is assumed.}

\begin{table}
    \centering
    \resizebox{0.9\columnwidth}{!}{%
    \begin{tabular}{l|ccc}
       $\mathbf{\Theta}$ & $\pi(\mathbf{\Theta},\mathcal{H}_1)$ & $\pi(\mathbf{\Theta},\mathcal{H}_0)$ & injection values \\
        \hline
       $p\ [M_\bullet]$ & $ \mathcal{U}[50, 1000]$ & & 100 \\ 
       $e$ & $\mathcal{U}[0, 0.9]$ & & 0.04\\
       $\cos(\theta_{\rm min})$ & $\mathcal{U}[0, 1]$ & & 0.5 \\
       $q_{r,{\rm ini}}$ & $\mathcal{U}[0, 2\pi]$ & & $1.4\pi$\\
       $q_{z,{\rm ini}}$ &$\mathcal{U}[0, 2\pi]$ & & $1.6\pi$\\
       $q_{\phi,{\rm ini}}$ & $\mathcal{U}[0, 2\pi]$ & & $0.3\pi$ \\
       $T_{\rm obt}\ [{\rm ks}]$ & $\mathcal{U}[20, 40]$ & & 31.5\\
       $a$ & $\mathcal{U}[0, 1]$ & & 0.9\\
       $\theta_{\rm obs}$ & $\mathcal{U}[0, \pi]$ & $\mathcal{U}[0, \pi/2]$ & $\pi/3$\\
       $\alpha_{\rm ini}$ & $\mathcal{U}[0, 2\pi]$ & None & 0\\
       $\tau_p\ [{\rm days}]$ & $\mathcal{U}[1, 2000]$ & None & 7 \& 20 \& 150\\
       $\beta_{\rm ini}$ & $\mathcal{U}[0, \pi/2]$ & None & $\pi/7$\\
       $\tau_a\ [{\rm days}]$ & $\mathcal{U}[1, 2000]$ & None & 500\\
       $\dot{T}_{\rm obt} \ [\times10^{-5}]$ & None & $\mathcal{U}[-10, 0]$ & -1\\
       $\dot{T}_{\rm obt,max} \ [\times10^{-5}]$ & $\mathcal{U}[-10, 0]$ & None & -1\\
       $\sigma_{\rm sys} \ [{\rm ks}]$ & $\mathcal{U}[0, 2]$ & & 0
    \end{tabular} }
    \caption{Priors and the injection values of the QPE timing model parameters used for the simulations. Entries left blank in the $\pi(\mathbf{\Theta}, \mathcal{H}_0)$ column indicate that the same prior values from the $\pi(\mathbf{\Theta}, \mathcal{H}_1)$ column are imposed.
    } 
    \label{tab:stra_pri}
\end{table}

Different from traditional X-ray telescopes, CATCH consists of $\mathcal{O}(10^2)$ small satellites which enable uninterrupted long timescale monitoring of multiple targets.
A month long continuous monitoring of a number of particularly interesting sources is practical for CATCH \citep{catch}.
As a reference, we first consider an uninterrupted observation epoch lasting for one apsidal precession period $T_{\rm aps}\approx 12$ days,
then two different strategies of observation cadence and duration of each observation epoch if the total amount of observation time is fixed as $T_{\rm obs}=4$ days.

\begin{figure*}
\includegraphics[scale=0.48]{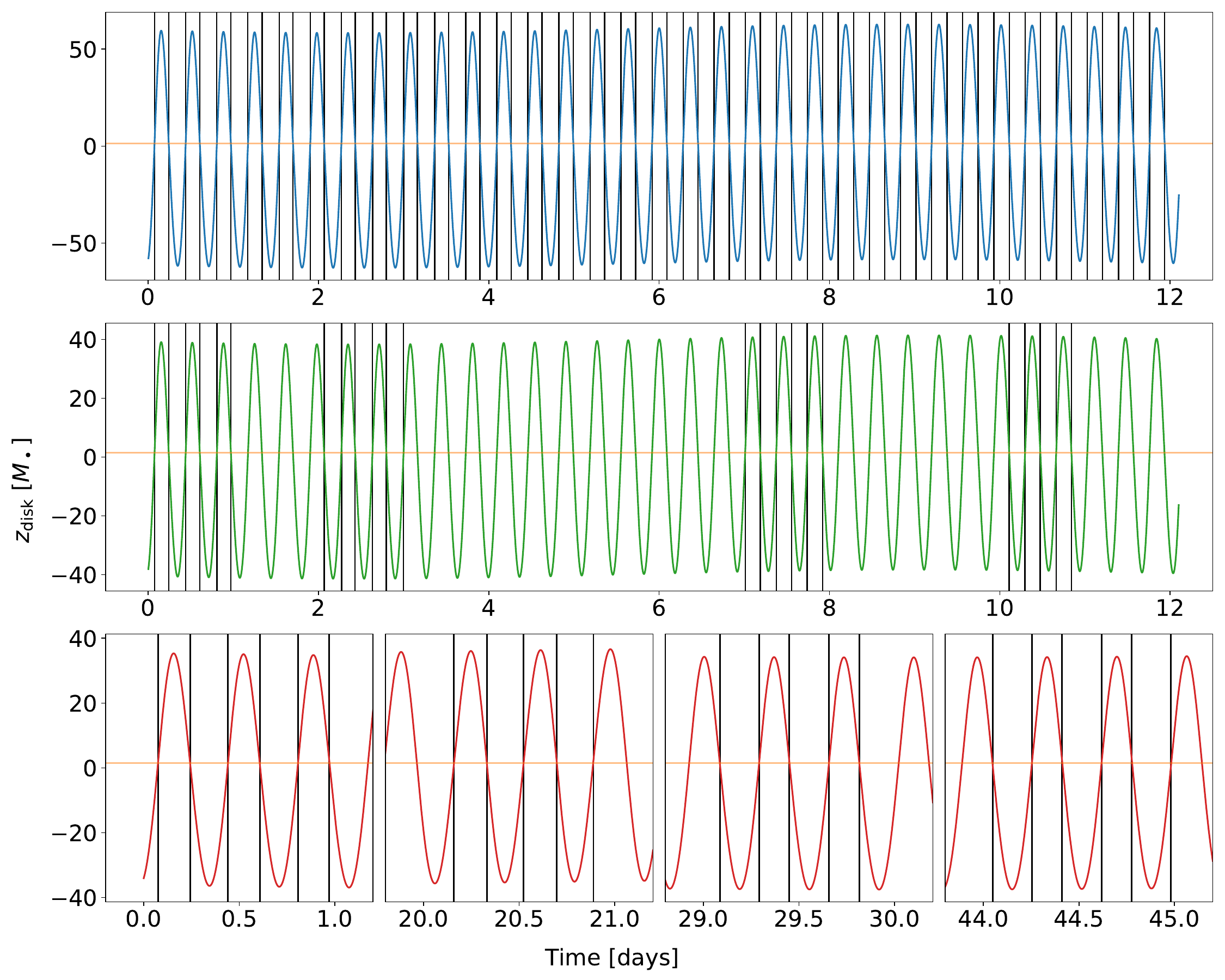}
\caption{\label{fig:mass_traj} Best-fit EMRI trajectories obtained with  three different strategies, where $z_{\rm disk}(t)$ is the $z$-component. The vertical bands indicate the simulated data $t_0^{(k)}\pm\sigma(t_0^{(k)})$. The orange horizontal line marks the disk surface. Blue: strategy A. Green: strategy B. Red: strategy C. 
}
\end{figure*}

\begin{description} 
    \item[Strategy A:]A single uninterrupted observation epoch lasting for one apsidal precession period $T_{\rm aps}\approx 12$ days.
    \item[Strategy B:]Four observation epochs distributed randomly in $12$ days, with each epoch lasting for 1 day.
    \item[Strategy C:]Same to B, except for the epochs distributed randomly in $\sim 50$ days, mimicking existing QPE observations, e.g.,  observations of GSN 069 during Dec. 2018-May 2019 (Fig.~\ref{fig:lcs}).
\end{description}


From mock observations of Strategies A, B and C, we infer model parameters as in the case of GSN 069.
The model parameter priors used are listed in Table~\ref{tab:stra_pri}. 
The inferred values of the semi-latus rectum $p$, eccentricity $e$, orbital period $T_{\rm obt}$, apsidal precession period $T_{\rm aps}$ and SMBH mass $M_\bullet$ from the three strategies are summarized in Table~\ref{tab:mass_res}.
The best-fit EMRI trajectories under the three strategies are shown in Fig.~\ref{fig:mass_traj}.

\begin{figure}
\includegraphics[scale=0.25]{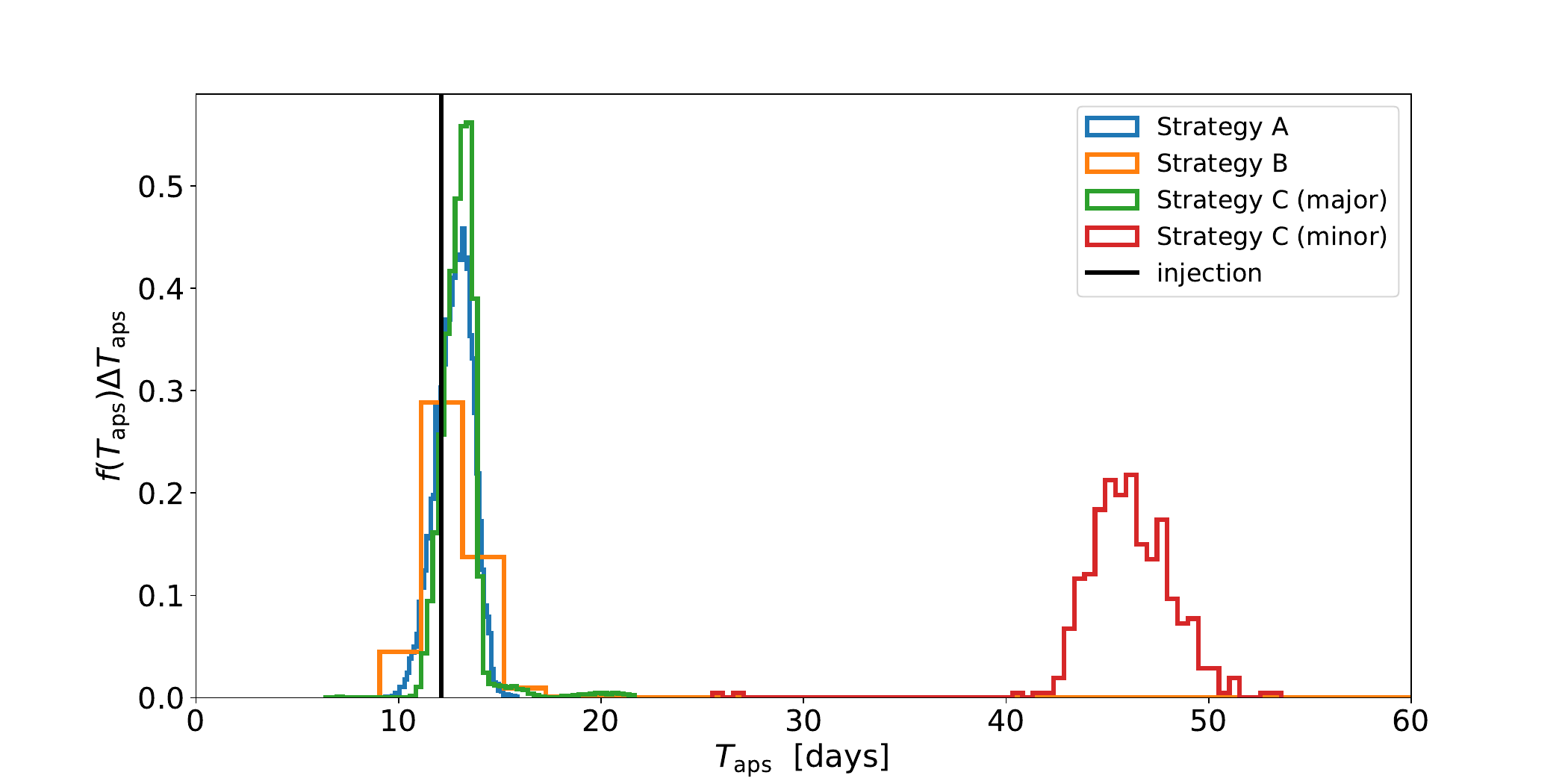}
\caption{\label{fig:T_aps} Probability distributions $f(T_{\rm aps})\Delta T_{\rm aps}$ of the apsidal precession period $T_{\rm aps}$ obtained 
from three different observation strategies, where $\Delta T_{\rm aps}$ is the bin size. 
}
\end{figure}

\begin{table*}
    \centering
    \resizebox{2\columnwidth}{!}{%
    \begin{tabular}{l|ccccc}
       Strategy for mass measurement & $p\ [M_\bullet]$ & $e$ & $T_{\rm obt}\ [{\rm ks}]$ & $T_{\rm aps}\ [{\rm day}]$ & $\log_{10}(M_\bullet/M_\odot)$ \\
        \hline
       A & $106^{+12}_{-16}$ & $0.04^{+0.00}_{-0.00}$ & $31.5^{+0.03}_{-0.06}$ & $12.9^{+1.5}_{-2.0}$ & $5.96^{+0.11}_{-0.07}$ \\
       B & $104^{+29}_{-18}$ & $0.04^{+0.01}_{-0.01}$ & $31.5^{+0.04}_{-0.07}$ & $12.6^{+3.5}_{-2.2}$ & $5.98^{+0.12}_{-0.16}$\\
       C (major/minor) & $107^{+19}_{-13}/378^{+34}_{-26}$ & $0.05^{+0.02}_{-0.02}$ & $31.5^{+0.03}_{-0.06}$ & $13.1^{+2.4}_{-1.6}/46.0^{+4.1}_{-3.1}$ & $5.95^{+0.08}_{-0.11}/5.13^{+0.05}_{-0.06}$ \\ 
    \end{tabular} }
    \caption{Median values and 2-$\sigma$ uncertainties of semi-latus rectum $p$, eccentricity $e$, orbital period $T_{\rm obt}$, apsidal precession period $T_{\rm aps}$ and SMBH mass $M_\bullet$ for the three strategies considered.} 
    \label{tab:mass_res}
\end{table*}

Specifically, we show the posteriors of the apsidal precession period $T_{\rm aps}$ in Fig.~\ref{fig:T_aps}. 
It is of no surprise to find uninterrupted monitoring for a long timescale ($\sim 12$ days) in Strategy A 
yields the tightest constraints on $T_{\rm aps}$ and consequently on the SMBH mass $M_\bullet$.
If the total observation time is limited, observation cadence and duration
of each observation epoch make a difference. 
Strategy B and C are similar except different observation cadence, which yield quite different $T_{\rm aps}$ constraints. 
Strategy B of high observation cadence also yields a reasonable constraint, wider than in Strategy A by a factor of $\sim \sqrt{3}$. 
In contrast, Strategy C of low observation cadence yields double-peak structure in the posterior of $T_{\rm aps}$, 
a major peak located at the injection value and a minor one at roughly four times the injection value, in addition to two extra tiny peaks at twice and three times the injection value, respectively.
For each peak, the fractional uncertainty $\delta T_{\rm aps}/T_{\rm aps}$ is comparable to in Strategy B, due to comparable total observation time.
This result is also consistent with the intuition that the phase information encoded in each observation epoch adds up “coherently" if intervals between consecutive observations is 
short ($\ll T_{\rm aps}$) as in Strategy B. Otherwise, multiple peaks  in the posterior may show up as in Strategy C. \footnote{Multiple peaks in the posterior may pose a challenge for some samplers. The sampler \texttt{nessai} with a large number of live points used in this work seem to properly sample all the peaks (see Fig.~\ref{fig:T_aps} for example).} 

For the purpose of precise measurement of SMBH masses from QPE timing, the optimal strategy is clearly strategy A, which consists of a single observation epoch that spans an entire apsidal precession period $T_{\rm aps}$. However, QPE sources exhibit a wide range of $T_{\rm aps}$ from tens to hundreds of of days. For sources of particular interest, such as eRO-QPE2 with $T_{\rm aps}\sim1$ month, uninterrupted long-term monitoring over one full $T_{\rm aps}$ is feasible for CATCH. Nevertheless, this approach becomes impractical for sources with longer $T_{\rm aps}$. A more practical alternative exemplified by strategy B is to divide the limited total observation time into $N\approx T_{\rm obs}/(2T_{\rm obt})$ separate epochs, ensuring that each epoch captures at least $\gtrsim3$ QPE flares. To properly sample the phase information of apsidal precession, these epochs should be quasi-uniformly distributed within one $T_{\rm aps}$, as exemplified in Strategy B.

\subsection{Disk motion}\label{subsec:model robustness}
In the previous subsection, we have applied the  QPE timing method to mock QPE data assuming an equatorial disk ($\mathcal{H}_0$).
As elaborated in Section~\ref{subsec:disk motion}, the QPE timing method is flexible in incorporating the component of disk motion ($\mathcal{H}_1$).
Both hypotheses can be adopted in analyzing the QPE timing data and the support to one hypothesis to another by data can be  quantified in three ways as shown in Section~\ref{subsec:gsn069}:
the best-fit models in the two hypotheses, the Bayes factor $\mathcal{B}_0^1$, and the constraint of disk precession period $\tau_{\rm p}$ in  $\mathcal{H}_1$.
For demonstrating the robustness of the QPE timing method for more general cases where the disk is precessing and aligning,
we generate three more sets of mock QPE data assuming a $\sim 12$ day long observation of an EMRI+disk system with same disk motion parameters $\alpha_{\rm ini}=0,\ \beta_{\rm ini}=\pi/7, \ \tau_a=500$ days and different disk precession periods $\tau_{\rm p}=7$ days, $20$ days and $150$ days, respectively.
All other injection parameters remain the same as in the previous subsection (Table~\ref{tab:stra_pri}). For comparison, we also generate a set mock data assuming an equatorial disk as in previous subsection.

\begin{table*}
    \centering
    \resizebox{2\columnwidth}{!}{%
    \begin{tabular}{c|ccccccc c}
       Injection & Hypothesis & $p\ [M_\bullet]$ & $e$ & $T_{\rm obt}\ [{\rm ks}]$ & $T_{\rm aps}\ [{\rm day}]$ &$\tau_{\rm p}$ [day] & $\log_{10}(M_\bullet/M_\odot)$  & $\log\mathcal{B}_0^1$\\
        \hline
        Precessing  disk
       & $\mathcal{H}_0$ & $604^{+376}_{-471}$ & $0.08^{+0.16}_{-0.07}$ & $31.6^{+0.06}_{-0.06}$ & $73.6^{+45.8}_{-57.4}$ & & $4.82^{+0.99}_{-0.32}$ & \\
        ($\tau_{\rm p} = 7$ d) & $\mathcal{H}_1$ & $107^{+13}_{-10}$ & $0.04^{+0.01}_{-0.01}$ & $31.5^{+0.04}_{-0.03}$ & $13.0^{+1.6}_{-1.2}$ & $7.1^{+0.1}_{-0.1}$ & $5.95^{+0.06}_{-0.07}$ & $213.2\pm0.2$\\
        \hline
       & $\mathcal{H}_0$ & $603^{+379}_{-490}$ & $0.06^{+0.19}_{-0.06}$ & $31.4^{+0.06}_{-0.06}$ & $73.0^{+45.9}_{-59.5}$ & &$4.82^{+1.10}_{-0.32}$ &  \\ 
        ($\tau_{\rm p} = 20$ d)& $\mathcal{H}_1$ & $112^{+44}_{-24}$ & $0.05^{+0.02}_{-0.02}$ & $31.5^{+0.17}_{-0.14}$ & $13.6^{+5.5}_{-3.0}$ & $19.8^{+21.7}_{-5.7}$ & $5.93^{+0.16}_{-0.22}$ & $194.0\pm0.2$\\
        \hline
       & $\mathcal{H}_0$ & $108^{+20}_{-17}$ & $0.04^{+0.01}_{-0.00}$ & $31.5^{+0.03}_{-0.06}$ & $13.2^{+2.5}_{-2.1}$ & &$5.94^{+0.11}_{-0.11}$ &  \\ 
        ($\tau_{\rm p} = 150$ d)& $\mathcal{H}_1$ & $109^{+18}_{-12}$ & $0.04^{+0.01}_{-0.01}$ & $31.5^{+0.02}_{-0.02}$ & $13.3^{+2.2}_{-1.5}$ & $1250^{+704}_{-857}$ & $5.94^{+0.08}_{-0.10}$ & $0.4\pm0.2$ \\
          \hline
        Equatorial disk
       & $\mathcal{H}_0$ & same as the Strategy A in Table~\ref{tab:mass_res}\\
        ($\tau_{\rm p} = \infty$ ) & $\mathcal{H}_1$ & $106^{+11}_{-11}$ & $0.04^{+0.01}_{-0.01}$ & $31.5^{+0.02}_{-0.03}$ & $12.9^{+1.3}_{-1.3}$ & $1181^{+770}_{-849}$ & $5.96^{+0.07}_{-0.06}$ & $-1.2\pm0.2$\\
    \end{tabular} }
    \caption{Median values and 2-$\sigma$ uncertainties of semi-latus rectum $p$, eccentricity $e$, orbital period $T_{\rm obt}$, apsidal precession period $T_{\rm aps}$, disk precession period $\tau_{\rm p}$ and SMBH mass $M_\bullet$ for different injections and hypotheses. } 
    \label{tab:robust_test}
\end{table*}

We fit each set of mock data with the EMRI+disk model with hypotheses $\mathcal{H}_0$ (an equatorial disk) and $\mathcal{H}_1$ (a precessing disk), respectively. The priors used in the Bayesian inference are listed in Table~\ref{tab:stra_pri} and the constraints on a fraction of model parameters 
obtained are summarized in Table~\ref{tab:robust_test}. We summarize the analysis results as follows.

\begin{description}
    \item[Fast disk precession with $\tau_{\rm p} = 7$ d or $20$ d:]The best-fit EMRI trajectories with the two hypotheses $\mathcal{H}_{0,1}$ are shown in Fig.~\ref{fig:robust_t7} and Fig.~\ref{fig:robust_t20}, 
from which, the disk-precessing hypothesis $\mathcal{H}_1$ is clearly favored over  the equatorial disk hypothesis $\mathcal{H}_0$.
This intuition is also supported  by large log Bayes factors $\log \mathcal{B}^1_0$ summarized in Table~\ref{tab:robust_test}. With the data favored hypothesis $\mathcal{H}_1$, it is of no surprise to find that all the physical parameters, including the semi-latus rectum $p$, eccentricity $e$, orbital period $T_{\rm obt}$, apsidal precession period $T_{\rm aps}$, SMBH mass $M_\bullet$ and the disk precession period $\tau_{\rm p}$ are correctly recovered and tightly constrained (Table~\ref{tab:robust_test}).
The other hypothesis $\mathcal{H}_0$ with a conflicting assumption is strongly disfavored by the mock data according to both the not well-behaved best-fit trajectory and 
the log Bayes factor. \\ 
With  hypothesis $\mathcal{H}_0$, the EMRI orbital period $T_{\rm obt}$ is still tightly constrained since they are mostly constrained by the alternating long-short pattern in the recurrence times,
therefore is not strongly affected by the model assumption on the disk motion, 
while the apsidal precession alone cannot properly fit two different super-orbital modulations introduced by both the apsidal precession and the disk precession,
therefore the apsidal precession period $T_{\rm aps} (\propto p) $ is not well constrained. 
\item[Slow disk precession with $\tau_{\rm p} = 150$ d:]The best-fit EMRI trajectories with the two hypotheses $\mathcal{H}_{0,1}$ are shown in Fig.~\ref{fig:robust_t150}.
The two best-fit trajectories are quite similar, because the slow disk precession makes little difference to the QPE timing during the relatively short observation span. Consequently, the disk precession period $\tau_{\rm p}$ in $\mathcal{H}_1$ is barely constrained, where no disk precession ($\tau_{\rm p}\rightarrow \infty$) is also compatible with the mock data, and the upper end of the 2-$\sigma$ credible interval obtained is in fact determined by the prior imposed.
Consistent with the resemblance of the two trajectories, the log Bayes factor between the two hypotheses $\log \mathcal{B}_0^1 = 0.4\pm 0.2$ also shows that there is no preference for one hypothesis over another from the mock data. 
As a result, the model parameter constraints in the two different hypotheses are also found to be similar. 

\item[Equatorial disk ($\tau_{\rm p}=\infty$):]The best-fit EMRI trajectories under the two hypotheses $\mathcal{H}_0$ and $\mathcal{H}_1$ are shown in Fig.~\ref{fig:robust_H0}. Similar to the slow precession case, the two trajectories are close due to the absence of disk-induced modulation. In hypothesis $\mathcal{H}_1$, the precession period $\tau_{\rm p}$ remains unconstrained, with the posterior dominated by the prior imposed. It is natural to find a log Bayes factor $\log \mathcal{B}_0^1 = -1.2 \pm 0.2$, which represents substantial preference for the true hypothesis $\mathcal{H}_0$. This case is quite similar to GSN 069.
\end{description}

The injection and recovery experiments demonstrate that EMRI apsidal precession and disk precession are not degenerate, both of which can be correctly recovered from QPE timing data, no matter the disk precession is fast, slow or absent.
In particular, for fast precessing disks ($\tau_{\rm p} \lesssim$ observation span), the disk precession period is  well constrained and the true hypothesis $\mathcal{H}_1$ is strongly favored according to the log Bayes factor $\log\mathcal{B}_0^1$. With a wrong hypothesis $\mathcal{H}_0$, the best-fit model prediction is clearly off the data simply because apsidal precession alone cannot explain modulations in QPE timing data from both apsidal precession and disk precession.  

For a slowly precessing disk ($\tau_{\rm p} \gg$ observation span), the disk precession period $\tau_{\rm p}$ in $\mathcal{H}_1$ is unconstrained, simply because disk-induced super-orbital modulations are negligible within the observation window. As a result, two best-fit trajectories in the two hypotheses are found to be quite similar, and 
there is no preference for equatorial disk precession hypothesis $\mathcal{H}_0$ over $\mathcal{H}_1$ according to the log Bayes factor. 

For an equatorial disk, the disk precession period $\tau_{\rm p}$ in $\mathcal{H}_1$ is again unconstrained. It is natural to find quite similar best-fit model predictions in the two hypotheses. The true hypothesis $\mathcal{H}_0$ is favored  according to the log Bayes factor.

\subsection{Optimal observation strategy: SMBH spin measurement}

Similar to the apsidal precession, Lense-Thirring precession of the EMRI orbital plane in the Kerr spacetime also introduces 
a super-orbital modulation to the QPE recurrence times. In general, the Lense-Thirring precession period $T_{\rm LT}$ is much longer than
both the orbital period $T_{\rm obt}$ and the apsidal precession period $T_{\rm aps}$. 
Uninterrupted monitoring of QPE sources for a long period of time $\sim T_{\rm LT}$ seems impractical even for CATCH.
To demonstrate the feasibility of the QPE timing method in measuring the SMBH spin 
with a limited amount of observation time $T_{\rm obs} \ll T_{\rm LT}$, we fix $T_{\rm obs} = 15$ days and simulate the following two observation
strategies.

\begin{description} 
    \item[Strategy I:]15 observation epochs with each epoch lasting for 1 day and distributed within $T_{\rm LT}$ in a quasi-uniform way.
    \item[Standard II:]5 observation epochs with each epoch lasting for 3 days and distributed within $T_{\rm LT}$ in a quasi-uniform way.
\end{description}

\begin{figure*}
\includegraphics[scale=0.28]{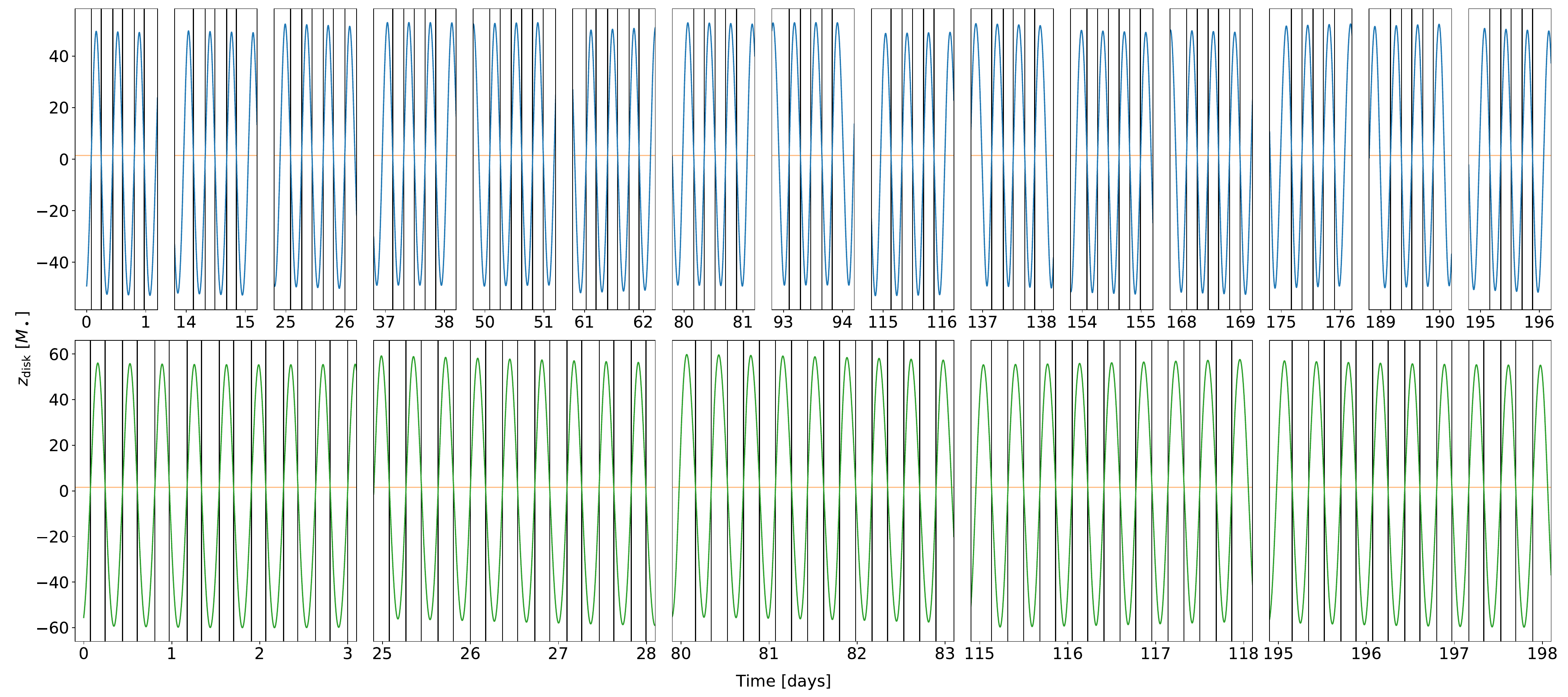}
\caption{\label{fig:spin_traj} Same to Fig.~\ref{fig:mass_traj} except for strategies to SMBH spin measurement. Blue: strategy I. Green: strategy II.
}
\end{figure*}

We perform the Bayesian analysis as in the case of SMBH mass measurement simulations.
The inferred values of the semi-latus rectum $p$, eccentricity $e$, orbital period $T_{\rm obt}$, apsidal precession period $T_{\rm aps}$, Lense-Thirring precession period $T_{\rm LT}$, SMBH mass $M_\bullet$ and SMBH spin $a$ from the two strategies are summarized in Table~\ref{tab:spin_res}.
The best-fit EMRI trajectories under the three strategies are shown in Fig.~\ref{fig:spin_traj}.

\begin{table*}
    \centering
    \resizebox{2\columnwidth}{!}{%
    \begin{tabular}{l|ccccccc}
       Strategy for spin measurement & $p\ [M_\bullet]$ & $e$ & $T_{\rm obt}\ [{\rm ks}]$ & $T_{\rm aps}\ [{\rm day}]$ & $T_{\rm LT}\ [{\rm day}]$ & $\log_{10}(M_\bullet/M_\odot)$ & $a$ \\
        \hline
       I & $100^{+1}_{-1}$ & $0.04^{+0.005}_{-0.005}$ & $31.5^{+0.005}_{-0.004}$ & $12.1^{+0.1}_{-0.1}$ & $203^{+22}_{-14}$ & $6.00^{+0.01}_{-0.01}$ & $0.89^{+0.07}_{-0.08}$ \\
       II & $101^{+3}_{-1}$ & $0.04^{+0.01}_{-0.01}$ & $31.5^{+0.01}_{-0.01}$ & $12.2^{+0.3}_{-0.2}$ & $213^{+24}_{-13}$ & $6.00^{+0.01}_{-0.02}$ & $0.86^{+0.06}_{-0.08}$ \\
    \end{tabular} }
    \caption{Median values and 2-$\sigma$ uncertainties of semi-latus rectum $p$, eccentricity $e$, orbital period $T_{\rm obt}$, apsidal precession period $T_{\rm aps}$, Lense-Thirring precession period $T_{\rm LT}$, SMBH mass $M_\bullet$ and SMBH spin $a$ for the two strategies considered.} 
    \label{tab:spin_res}
\end{table*}

Although observations of low  cadence are not able to resolve apsidal precession in strategy C, more observations as in strategy I, make a difference. If enough amount of separated observation epochs are distributed within one $T_{\rm LT}$, strategy I demonstrates that $T_{\rm LT}$ can be resolved, which in turn improves the resolution of $T_{\rm aps}$. Meanwhile, the strategy II with an even lower observation cadence yet longer single epoch duration within one $T_{\rm LT}$ achieves the same precision of $M_\bullet$ and $a$ measurement as in strategy I. Both strategy I and II reduce the 2-$\sigma$ uncertainty of $\log_{10}(M_\bullet)$ by an order of magnitude compared to  strategies A-C. The SMBH spin $a$ is also well constrained with comparable 2-$\sigma$ uncertainties $<0.1$ in both strategies I and II. 
Strategy I with higher cadence yields tighter constraints on a fraction of model parameters $(e, T_{\rm obt}, T_{\rm aps})$ than in strategy II.

A similar idea has also been proposed for measuring the Sgr A$^\star$ SMBH spin by tracing S stars \citep{Zhang:2015nya}.
For the star S2, the Lense–Thirring precession is out of reach of the current astrometry \citep{Genzel:2024vou}.
A star moving around the Sgr A$^\star$ on a tighter orbit is desired for detecting 
the SMBH spin from Lense–Thirring precession \citep{waisberg2018}.


\section{Summary and Discussions}\label{sec:conclusion}
More and more analyses show that (at least a  fraction of) QPEs are sourced by quasi-periodic collisions between SMOs and  accretion disks of  SMBHs.
As a result, the QPE timing data encode rich information of the EMRI trajectory, including the orbital period $T_{\rm obt}$, the apsidal precession period $T_{\rm aps}$
and the Lense-Thirring precession period $T_{\rm LT}$, If these periods are extracted from QPE timing data, one then can robustly measure the SMBH mass and spin. We name this dynamical measurement method as the QPE timing method,
which is in the same spirit of measuring the Sagittarius A$^\star$ SMBH mass by tracing individual stars around the SMBH ~\citep{Schodel2002,Ghez2005,Gillessen2009,GRAVITY2018b}. 
The QPE timing method can be applied to a much broader range of SMBHs  since there is no need to spatially resolve the orbits of individual stars.

To implement the QPE timing method, we constructed a comprehensive Bayesian inference framework in which one can reconstruct the EMRI trajectory (and the disk motion), 
consequently measure the SMBH parameters from QPE timing data.

As an example, we first applied the QPE timing method to the well-studied QPE source GSN 069, analyzing its orbital parameters under different hypotheses in the Bayesian inference framework.
In the disk precession hypothesis $\mathcal{H}_1$, the disk precession period $\tau_{\rm p}$ is found to be longer than the observation span, therefore introduces negligible modulation to the QPE timing. This observation is also supported by similar best-fit EMRI trajectories in $\mathcal{H}_1$ and in $\mathcal{H}_0$ (an equatorial disk). Consistent with this observation, 
the log Bayes factor [Eq.~(\ref{eq:logB})] between two different hypotheses also represents substantial preference for $\mathcal{H}_0$.
In the data favored hypothesis $\mathcal{H}_0$, the SMBH mass is tightly constrained [Eq.~(\ref{eq:GSN069_mass})], with an uncertainty lower than intrinsic uncertainties of $M_\bullet-\sigma_\star$ relations. The other hypothesis $\mathcal{H}_1$ yields similar model parameter constraints.

To further demonstrate the robustness of the QPE timing method in the presence of disk motion, we generated additional mock QPE datasets assuming a precessing and aligning disk with a wide range of precession periods, and analyzed them under both hypotheses $\mathcal{H}_0$ and $\mathcal{H}_1$.
As a result, we find that $\mathcal{H}_1$ is strongly favored, in terms of Bayes factor, trajectory best-fits and the constraint of disk precession period $\tau_{\rm p}$, when the disk is precessing rapidly. In the case of a slowly precessing disk with $\tau_{\rm p}$ much longer than the observation span, the disk precession period $\tau_{\rm p}$ is unconstrained;  
neither hypothesis is strongly preferred according to the log Bayes factor; however, both yield similar accurate recovery of EMRI orbital parameters. 
In the case of an equatorial disk, the disk precession period $\tau_{\rm p}$ is unconstrained and the precessing disk hypothesis $\mathcal{H}_1$ is disfavored according to the log Bayes factor; EMRI orbital parameters are 
correctly recovered in both hypotheses.
Notably, these tests confirm that EMRI motion and disk motion are not degenerate in QPE timing data: the underlying EMRI dynamics can  be reliably inferred no matter the disk precession 
is fast, slow or absent
These examples demonstrate the feasibility of the QPE timing method in reconstructing the EMRI trajectory and the disk motion.

The QPE timing method opens up the possibility for precision measurement of SMBHs, though the current bottleneck is the limited  X-ray observation resources available.
The multi-target X-ray telescope under construction, CATCH \citep{catch}, will hopefully be the solution to the bottleneck of X-ray followup observation. First, CATCH will allow comprehensive X-ray monitoring of a large sample of optical TDEs in the era of the Rubin Observatory Legacy Survey of Space and Time (LSST, \citealt{LSST2019}) and the Wide Field Survey Telescope (WFST, \citealt{Wang2023WFST}), and thus the discovery of a large number of QPEs following TDEs. In fact, it is impractical to follow all TDEs that will be discovered in the near future even for CATCH. This is not necessary either. There is emerging evidence suggesting that QPEs are not equally associated with all TDEs, yet they occur at a much higher rate in recently faded AGNs~\citep{Jiang2025}. This preferred subset of TDEs can be identified by their bright infrared echoes from torus remnant~\citep{Wu2025} or by the presence of extened emission-line regions in their host galaxies~\citep{Xiong2025}. Accurate measurements of even a small number of SMBHs with QPE timing data are valuable, e.g.,
in calibrating traditional mass measurement methods in the low mass range.
In addition, different from traditional X-ray telescopes, CATCH consists of $\sim 10^2$ small satellites, a small number of which can be fully reserved for  TDE follow-up and QPE monitoring.
Refined observations can then be made for the QPE targets. 

In this work, we explored different observation strategies for accurately measuring SMBH masses from QPE timing in the CATCH era.
In general, the apsidal precession period $T_{\rm aps}$ is much longer than the orbital period $T_{\rm obt}$, therefore is harder to be accurately measured.
As a result, the mass $M_\bullet$ measurement precision is mainly determined
by the measurement precision  of the apsidal precession period, with 
\be 
\frac{\delta M_\bullet}{M_\bullet}\approx\frac{3}{2}\frac{\delta T_{\rm aps}}{T_{\rm aps}}\ .
\ee 
For particularly interesting QPE sources, the design of CATCH enables uninterrupted long-term (say $\sim 1$ month) monitoring.
This will of course be the optimal strategy for precision measurement of SMBH masses. 
For general QPE sources, we may allocate less total amount of observation time $T_{\rm obs}$, and divide the total observation time into $N\approx T_{\rm obs}/(2 T_{\rm obt})$ epochs
that are distributed in a quasi-uniform way within one apsidal precession period $T_{\rm aps}$. 
In this way, each observation epoch guarantees to detect $\gtrsim 3$ eruptions, and the apsidal precession information encoded in different epochs adds up “coherently", 
therefore enables an optimal measurement of $T_{\rm aps}$ given a limited amount of total observation time.

In addition to apsidal precession of the EMRI, a misaligned and precessing disk is also possible to introduce super-orbital modulations to the QPE timing. In this case, the disk motion information may also be extracted from the QPE data as demonstrated via the injection and inference experiments in Section~\ref{subsec:disk motion}.

The same strategy works for measuring the SMBH spin using the QPE timing method. The only difference is that a even longer timescale $T_{\rm LT}$ matters, which takes more observation time to be measured. The spin measurement precision is up to the measurement precision of 
the Lense-Thirring precession period $T_{\rm LT}$ with 
\be 
\frac{\delta a}{a}\approx \frac{\delta T_{\rm LT}}{T_{\rm LT}}\ .
\ee 
In practice, the apsidal precession period $T_{\rm aps}$ and the Lense-Thirring precession period $T_{\rm LT}$ are unknown {\it a priori}. 
Therefore, the observation strategy and  the model parameter inference can only be updated iteratively.

\section*{Acknowledgements}
We thank the anonymous referee for helpful comments. We thank Yue Liao for helping making the schematic picture in Fig.~2. This work is supported by the Strategic Priority
Research Program of the Chinese Academy of Sciences
(XDB0550200), the National Natural Science Foundation
of China (grants 12192221,12393814) and the China Manned Space Project.

\section*{Data Availability}

ObsIDs of the three XMM-Newton observations are 0823680101, 0831790701, 0851180401. The Chandra observation (ObsID:22096) of GSN 069 at 2019-02-14 can be obtained by the Chandra X-ray Observatory, contained in the Chandra Data Collection (CDC) 368~\href{https://doi.org/10.25574/cdc.368}{doi:10.25574/cdc.368}.



\bibliographystyle{mnras}
\bibliography{eg} 




\appendix
\section{Posterior corner plots of GSN 069}
In the section, we include posterior corner plots of model parameters in hypotheses $\mathcal{H}_0$ and $\mathcal{H}_1$ for analyzing QPE timing data of GSN 069.

\begin{figure*}
\includegraphics[scale=0.25]{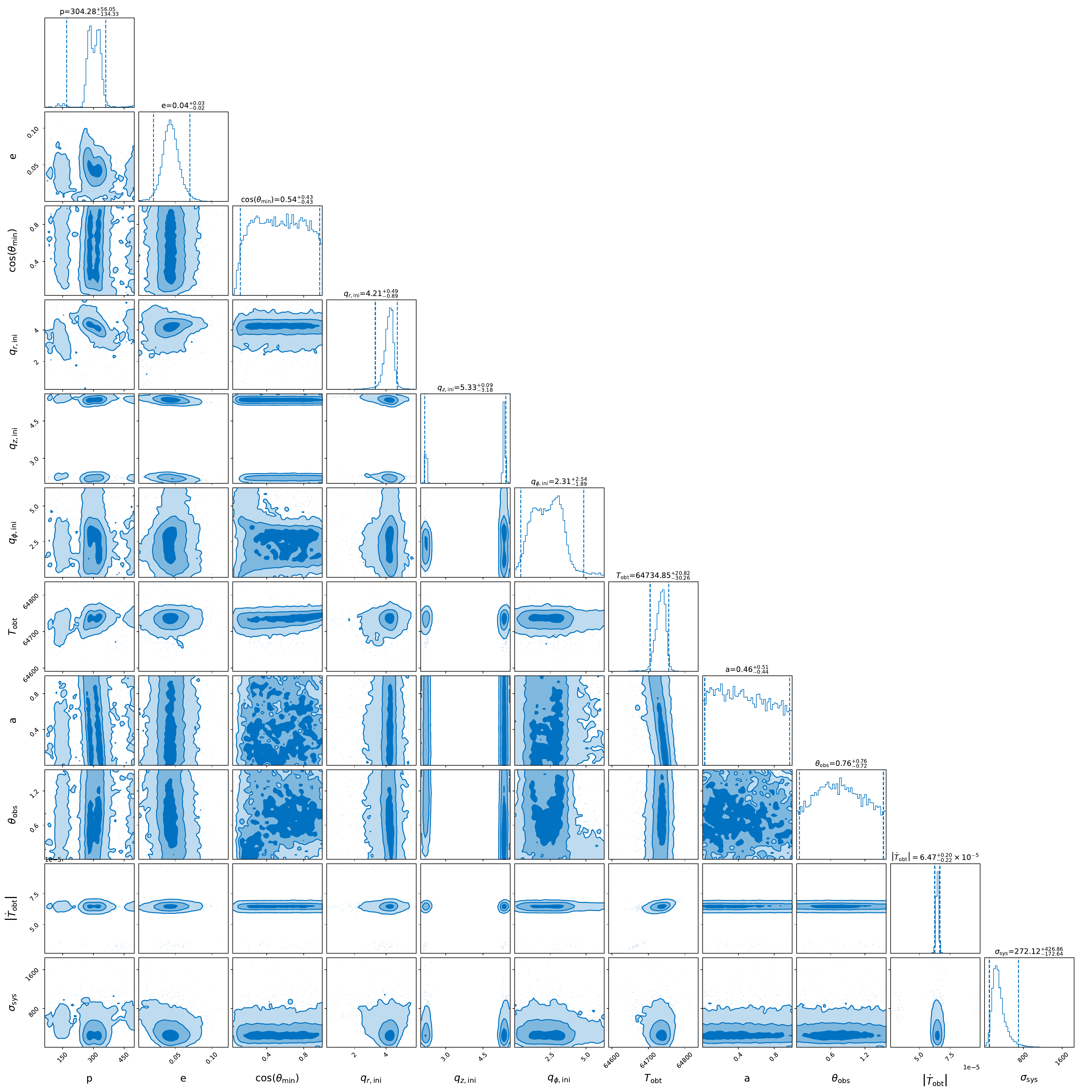}
\caption{\label{fig:gsn_h0}  The posterior corner plot of model parameters for GSN 069
 with the vanilla hypothesis($\mathcal{H}_0$): $p [M_\bullet], e, 
\cos\theta_{\rm min}, q_{r,\mathrm{ini}}, q_{z,\mathrm{ini}}, q_{\phi,\mathrm{ini}},
T_{\rm obt} [{\rm sec}], a, \theta_{\rm obs},  \dot{T}_{\rm obt} [\times10^{-5}], \sigma_{\rm sys} [{\rm sec}]$, where each pair of vertical lines denotes the 2-$\sigma$ confidence level.}
\end{figure*}

\begin{figure*}
\includegraphics[scale=0.24]{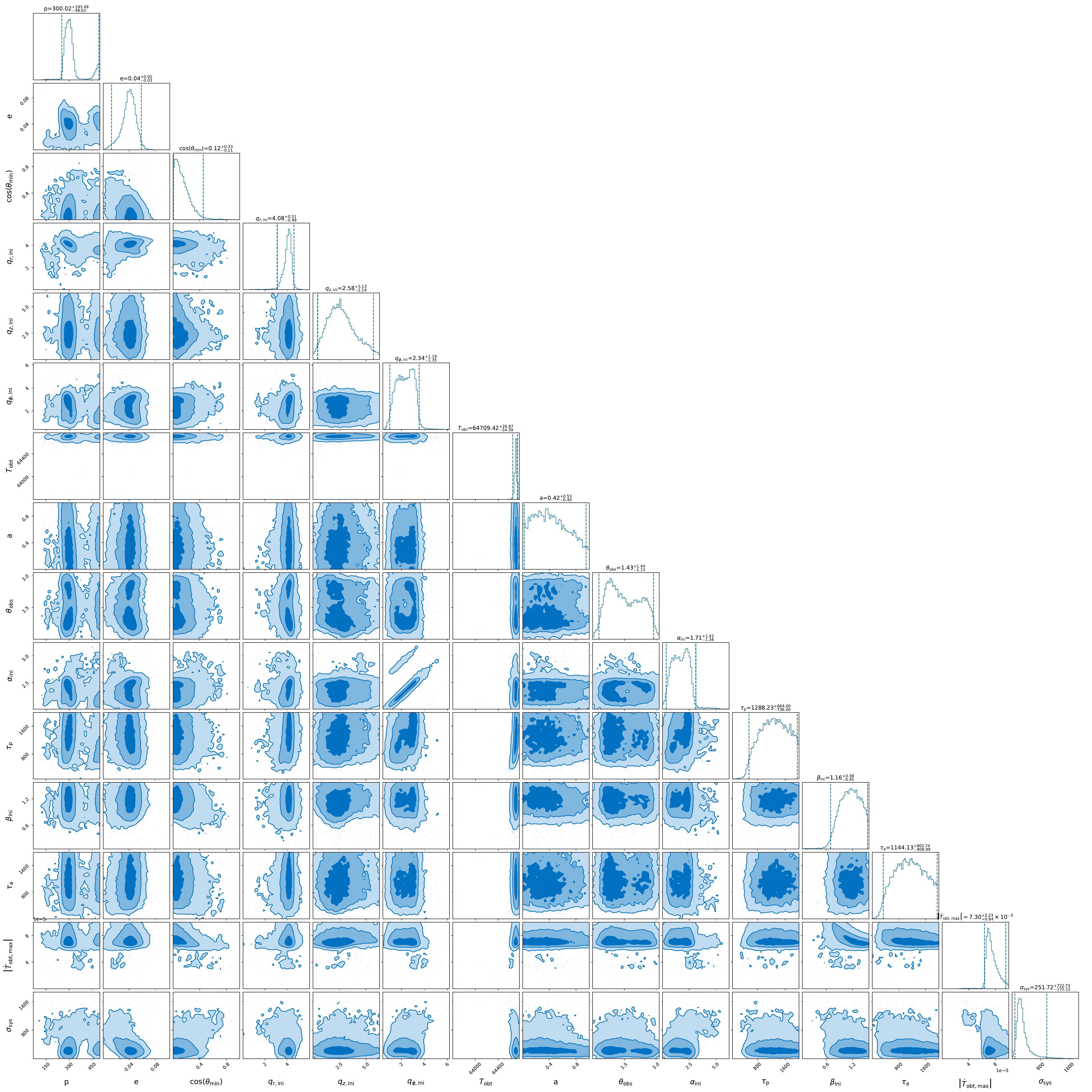}
\caption{\label{fig:gsn_h2}  The posterior corner plot of model parameters for GSN 069
 with the disk precession and alignment hypothesis($\mathcal{H}_1$): $p [M_\bullet], e, 
\cos\theta_{\rm min}, q_{r,\mathrm{ini}}, q_{z,\mathrm{ini}}, q_{\phi,\mathrm{ini}},
T_{\rm obt} [{\rm sec}], a, \theta_{\rm obs}, \alpha_{\rm ini}, \tau_{\rm p} [{\rm days}], \beta_{\rm ini}, \tau_{\rm a} [{\rm days}], \dot{T}_{\rm obt, max} [\times10^{-5}], \sigma_{\rm sys} [{\rm sec}]$, where each pair of vertical lines denotes the 2-$\sigma$ confidence level.}
\end{figure*}

\section{O-C analysis of GSN 069}\label{app_b}
\begin{figure*}
\includegraphics[scale=0.36]{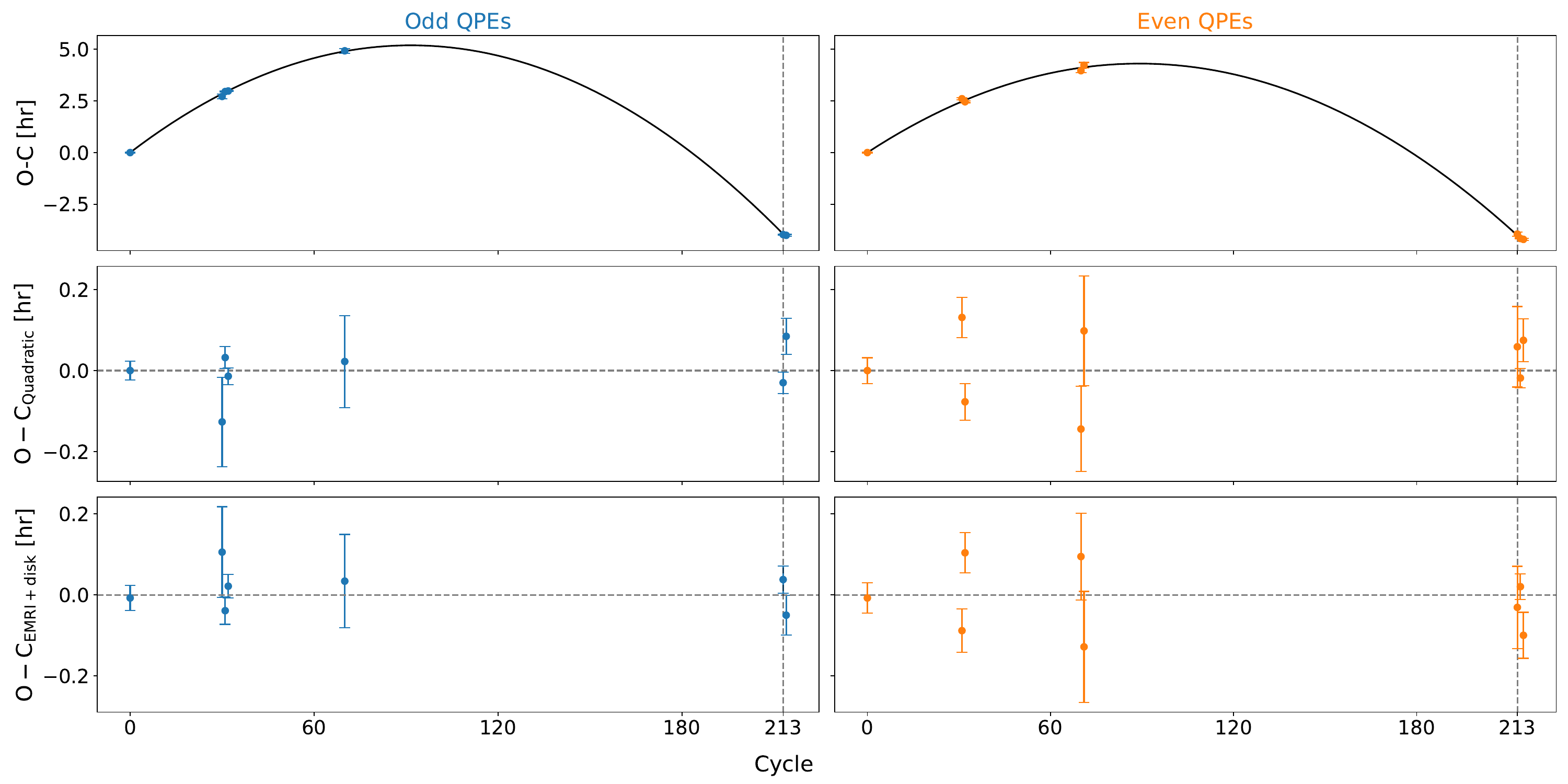}
\caption{\label{fig:gsn_oc} \bf O-C analysis versus EMRI+disk model fitting to GSN 069 data, where  error bars represent 1-$\sigma$ uncertainties. The number of cycles of the first flare in the last epoch is labeled by the dashed vertical line.}
\end{figure*}

For comparing with the Bayes analysis in the framework of EMRI+disk model, we apply a similar O-C analysis as in \cite{Miniutti2025} to GSN 069 data.
In the top two panels of Fig.~\ref{fig:gsn_oc}, we show the fitting result of a simple linear model, i.e., a constant period $T_{\rm trial}$,
\be 
{\rm O-C} = t_{\rm obs}^{(k)}-N_{\rm cyc}T_{\rm trial}\ ,
\ee 
where $N_{\rm cyc}$ is the number of cycles and $T_{\rm trial}=17.88$ hr \citep{Miniutti2025}. In the O-C analysis, there is some ambiguity in identifying $N_{\rm cyc}$ of each flare due to data gaps, while it is of no such difficulty in the Bayes analysis of the EMRI+disk model. Therefore, we use the numbers $N_{\rm cyc}$ from the best-fit EMRI+disk model in the O-C analysis.
Note that the first flare in the last epoch is identified as the 213th flare, while it was identified  as the 211th in \cite{Miniutti2025}.
From top panels of Fig.~\ref{fig:gsn_oc}, the linear model of a constant period is insufficient to fit the data. In middle panels, we use a quadratic model 
\be
    {\rm O-C_{quadratic}} = t^{(k)}_{\rm obs}-\left(T_{\rm obt}N_{\rm cyc}+\frac{1}{2}\dot{T}_{\rm obt}\langle T_{\rm obt}\rangle N_{\rm cyc}^2\right)\ ,
\ee
where $\langle T_{\rm obt}\rangle$ is the averaged orbital period across the whole time span assuming constant $\dot{T}_{\rm obt}$.
Fitting the quadratic model to either the odd and even flares, orbital period and its derivative are constrained as  
\be 
\begin{aligned}
    T_{\rm obt} &= 64776^{+4}_{-4} \ {\rm s}\ , \\
    \dot T_{\rm obt} &=-6.9^{+0.1}_{-0.1}\times10^{-5}\ ,\quad ({\rm Odd\ flares}) \\
\end{aligned}
\ee 
and 
\be 
\begin{aligned}
    T_{\rm obt} &= 64715^{+8}_{-8} \ {\rm s}\ , \\
    \dot T_{\rm obt} &=-6.1^{+0.1}_{-0.1}\times10^{-5}\ ,\quad ({\rm Even\ flares}) \\
\end{aligned}
\ee 
at 2-$\sigma$ confidence level. The O-C$_{\rm quadratic}$ analysis does not yield clear in-phase or anti-phase pattern in the residuals of even and odd flares. 
In the bottom panels are the residuals of the best-fit QPE timing model with $\mathcal{H}_0$ for GSN 069, i.e. $t^{(k)}_{\rm obs}-t_0^{(k)}$, 
in which there is no clear signature of unmodeled super-orbital modulation either. 

\begin{figure*}
\includegraphics[scale=0.36]{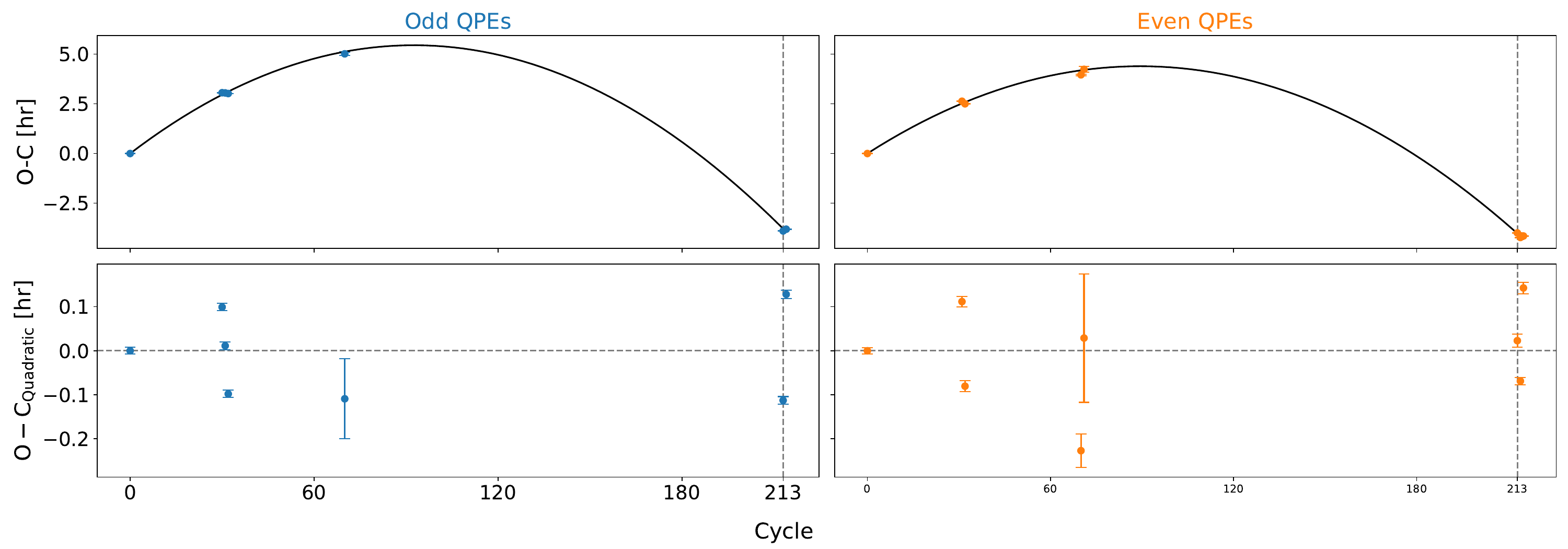}
\caption{\label{fig:gsn_oc_peak} \bf Same to Fig.~\ref{fig:gsn_oc} except for that flare peak times  instead of flare starting times are used.}
\end{figure*}

Another possible source of discrepancy is the different ways of defining the QPE timing: we define a starting time for each flare and interpret it as when the star crosses the disk, while \cite{Miniutti2025} use the flare peak time. We again apply the O-C analysis to flare peak times and the results are displayed in Fig.~\ref{fig:gsn_oc_peak}.  The results are quit similar to what is show in Fig.~\ref{fig:gsn_oc} except lower uncertainties of the flare peak times.

To summarize, there is some intrinsic ambiguity for identifying number of cycles of each flare in O-C analysis. This may lead to different interpretations of O-C analysis results and is likely the source of discrepancy between this work and \cite{Miniutti2025}.

\section{Best-fit EMRI trajectories  disk motion}
In this section, we present the best-fit EMRI trajectories with both $\mathcal{H}_0$ and $\mathcal{H}_1$ for the simulations in Section~\ref{subsec:model robustness}.

\begin{figure*}
\includegraphics[scale=0.48]{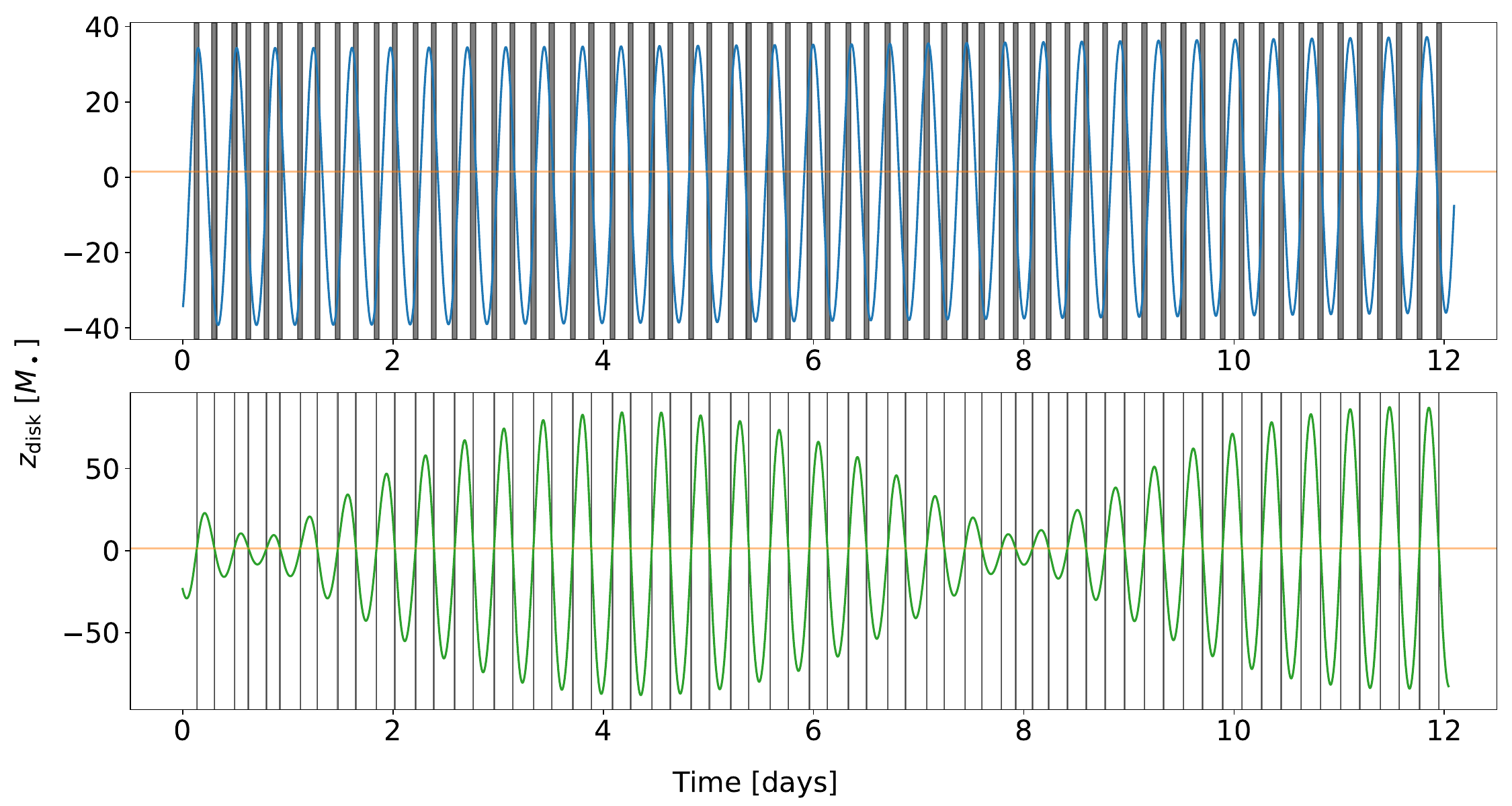}
\caption{\label{fig:robust_t7} Best-fit EMRI trajectories obtained with the two hypothesis in the fastest disk precession case with $\tau_{\rm p}=7$ days, where $z_{\rm disk}(t)$ is the distance to the disk midplane. The vertical bands indicate the simulated data with inference uncertainties $t_0^{(k)}\pm\tilde{\sigma}(t_0^{(k)})$, where $\tilde{\sigma}(t_0^{(k)})=\sqrt{(\sigma(t_0^{(k)}))^2+\sigma^2_{\rm sys}}$. The orange horizontal line marks the disk surface. Upper panel: vanilla hypothesis ($\mathcal{H}_0$). Lower panel: disk precession and alignment hypothesis ($\mathcal{H}_1$). It is clear that $\mathcal{H}_1$ is much more favored by the mock data.
}
\end{figure*}

\begin{figure*}
\includegraphics[scale=0.48]{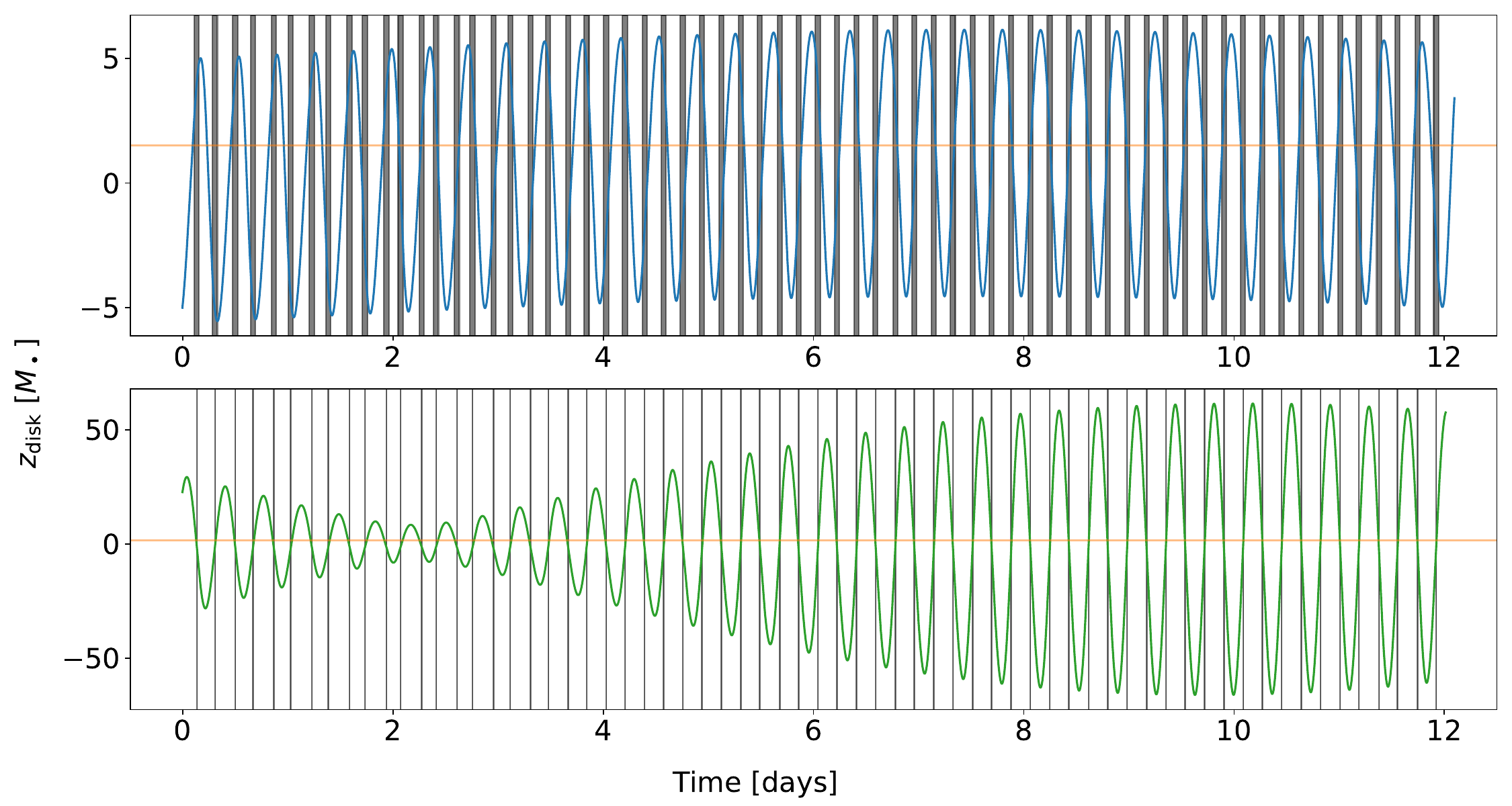}
\caption{\label{fig:robust_t20} Same to Fig.~\ref{fig:robust_t7} except for a precessing disk with $\tau_{\rm p}=20$ days.
}
\end{figure*}

\begin{figure*}
\includegraphics[scale=0.48]{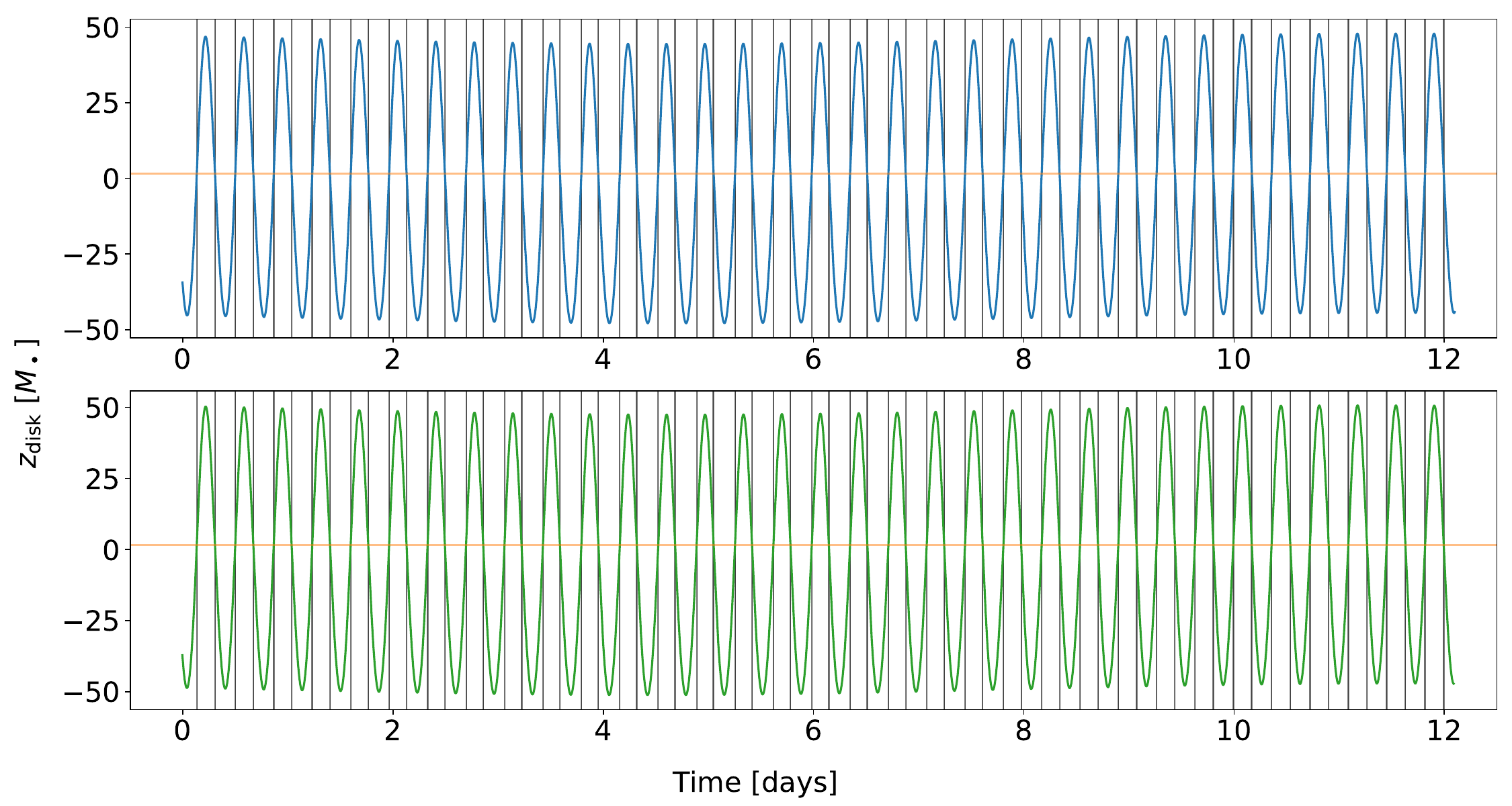}
\caption{\label{fig:robust_t150} Same to Fig.~\ref{fig:robust_t7} except for a slow precessing disk $\tau_{\rm p}=150$ days.
}
\end{figure*}

\begin{figure*}
\includegraphics[scale=0.48]{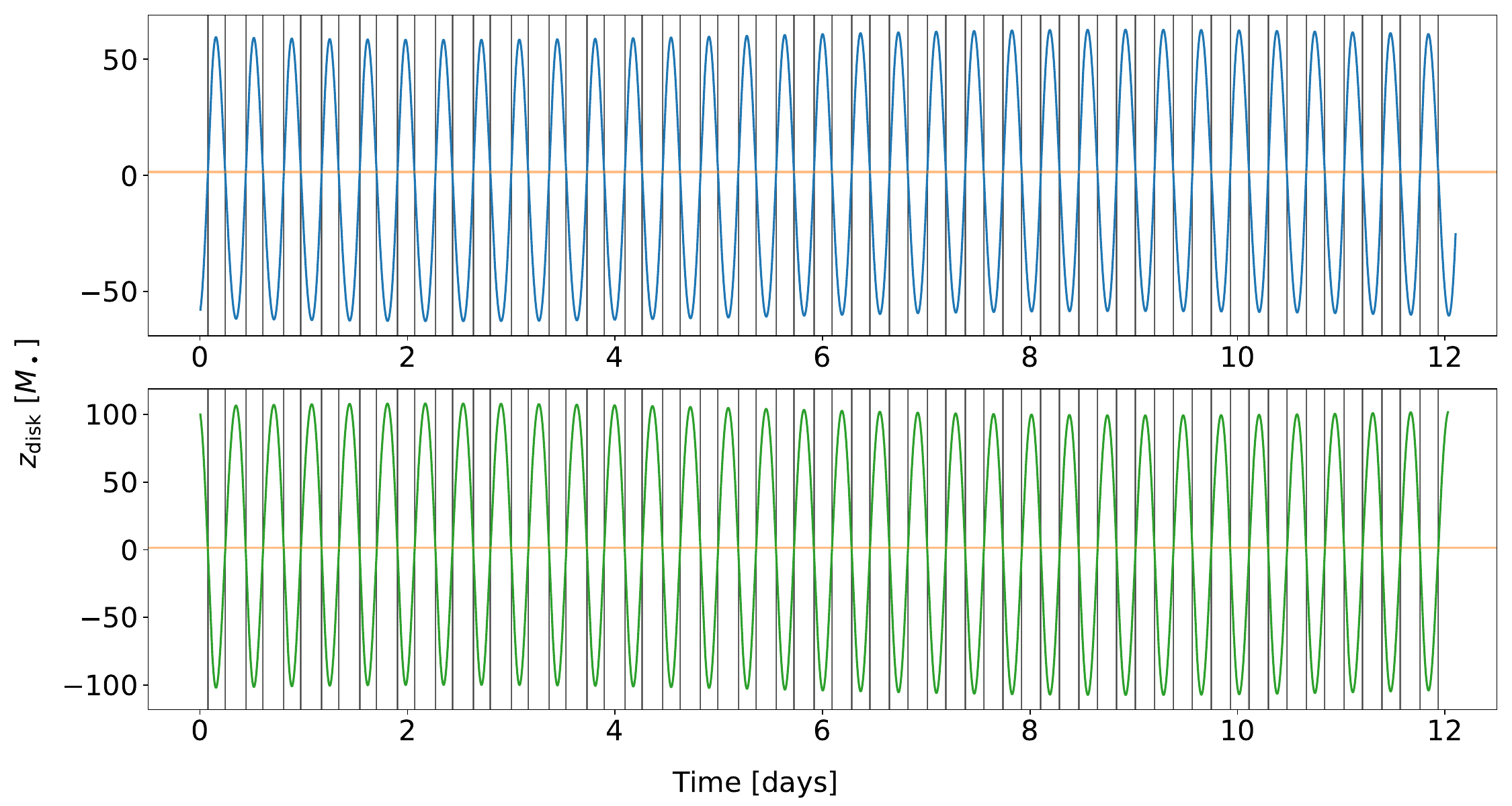}
\caption{\label{fig:robust_H0} Same to Fig.~\ref{fig:robust_t7} except for an equatorial disk $\tau_{\rm p}=\infty$ days. The upper panel is in fact the same as the upper panel in Fig.~\ref{fig:mass_traj}.
}
\end{figure*}



\bsp	
\label{lastpage}
\end{document}